\documentclass[reprint,amsmath,amssymb,aps]{revtex4-2}
\usepackage{tipa}
\usepackage{color}
\usepackage{bm,epsfig,mathrsfs,amsmath,amssymb,graphicx,subfigure,overpic}
\usepackage{graphicx}
\usepackage{dcolumn}
\usepackage{bm,ulem}

\usepackage{chngcntr}
\usepackage{float} 
\usepackage{booktabs}%
\newcommand{\PreserveBackslash}[1]{\let\temp=\\#1\let\\=\temp}
\newcolumntype{C}[1]{>{\PreserveBackslash\centering}p{#1}}
\newcolumntype{R}[1]{>{\PreserveBackslash\raggedleft}p{#1}}
\newcolumntype{L}[1]{>{\PreserveBackslash\raggedright}p{#1}}
\usepackage[colorlinks,citecolor=blue,linkcolor=red]{hyperref}
\usepackage[toc,title,titletoc,header]{appendix}

\date{\today}
\begin{document}
\title{The Filtering Demon: Beyond Standard Thermodynamic Bounds and Harnessing  Measurement Error and Quantum Friction}
\author{Yang Xiao$^{1}$}
\author{Jin Wang$^{2}$}\email{jin.wang.1@stonybrook.edu}
\affiliation{ $^1\,$College of Physics, Jilin University, Changchun 130022, China\\  
$^2\,$ Department of Chemistry and Department of Physics and Astronomy, State University of New York at Stony Brook, Stony Brook, New York 11794, USA}

\begin{abstract} 
Standard stochastic heat engines operate blindly,  enforcing work extraction protocols indiscriminately on microstates, and consequently suppressing work output, stability, efficiency.  To overcome this, we propose an Otto information engine (OIE) that employs a Maxwell’s demon to filter out detrimental stochastic trajectories,  and   demonstrate that the OIE can  provide enhanced work output and stability compared to the corresponding standard Otto engine. Remarkably, even after accounting for the energetic costs of the demon, the OIE efficiency surpasses both the standard Otto limit and Carnot bound. {Furthermore, by applying the fluctuation theorem of information dissipation, we derive upper and lower bounds on efficiency, confirming that our results adhere to the second law of thermodynamics.}
Finally, and counterintuitively,  measurement errors  and quantum inner friction, traditionally considered deleterious, can be  harnessed as resources, which  improves robustness and  enables us to ignore  the adiabatic strokes time.
\end{abstract}
\maketitle
\date{\today}

\section{Introduction}

Maxwell’s demon is a cornerstone of modern thermodynamics \cite{Maxwell1871,Vedral09}, providing a fundamental framework to explore the interplay between information and energy. Originally conceived to challenge the second law of thermodynamics \cite{Maxwell1871,Szilard1929,LB51,Landauer61,Bennett1982,Sagawa12,Zeng21}, the demon illustrates how an agent can rectify thermal fluctuations to extract work via measurement-based feedback \cite{JM19,SY16,LB19,Chen21,Ding18,JY96,
JJ13,CE17,CE18,Lutz23,SS20,KC17,JV15,LB21,Jar23,Jar12,TK21,TK23,JP23,
GP20,TV20}.  This principle is now recognized as a universal mechanism, governing phenomena ranging from biological molecular motors \cite{EM21,Wen18,Sc24,SF23,VS07} to quantum error correction \cite{BM15}. With recent advances in experiment technology, information-to-energy conversion has transitioned from a theoretical abstraction into a practical application, across diverse platforms, including Brownian systems \cite{BJ08, Sag10, GP18, YJ14}, electronic systems\cite{JV15, JV14, Sag14, KC17},  photonics \cite{MD16}, and superconducting circuits  \cite{NC17}.

Beyond its historical significance in macroscopic heat engines, the stochastic heat engines \cite{HT07,Lutz14,kosloff06,Xiao23,RJ19,JK17,Holubec18,Fei22,Seifert12,Wang24,DD23}  serve as a paradigmatic framework for exploring energy conversion at the nanoscale.  However, the stochastic heat engines  operate blindly, as external driving fields enforce compression or expansion without regard to the system's instantaneous microstate. This non-selective operation is inherently inefficient.
 For instance, adiabatically compressing a spin-1/2 system initially in its excited state requires work input rather than extraction. Within the framework of the stochastic thermodynamics, such unfavorable events not only diminish the net work output but also trigger large work fluctuations due to dispersed work distribution. This indiscriminate manipulation remains a fundamental barrier to the efficiency and stability of the stochastic heat engines.

Here, we overcome this thermodynamic blindness by introducing an Otto information engine (OIE) that  filters out the trajectories detrimental to work extraction by a demon. We derive exact analytical expressions for the work output, the work  fluctuations (reflect stability), and the efficiency of the OIE. Based on these, we demonstrate that the OIE outperforms its standard counterpart, delivering higher work output while  suppressing fluctuations to ensure stable work delivery. Remarkably, even when accounting for the energetic cost of the demon, the OIE efficiency surpasses both the standard Otto limit and  Carnot bound. {In addition, we obtain the upper and the low bounds of the OIE efficiency by the fluctuation theorem of information dissipation \cite{Zeng21}, which can be used to  demonstrate the rationality of our results.} Finally, we reveal a counterintuitive phenomenon:  measurement errors \cite{TK22,GP20,Sag14}  and quantum friction \cite{RJ19,JPS19,Plastina14} which are typically considered detrimental can be harnessed as valuable resources. This not only enhances the robustness of the engine, but also enables the OIE's adiabatic time to be ignored which can increase the  output power by  orders of magnitude.

\section{The standard Otto cycle}
\begin{figure}
    \centering
\includegraphics[width=1\linewidth]{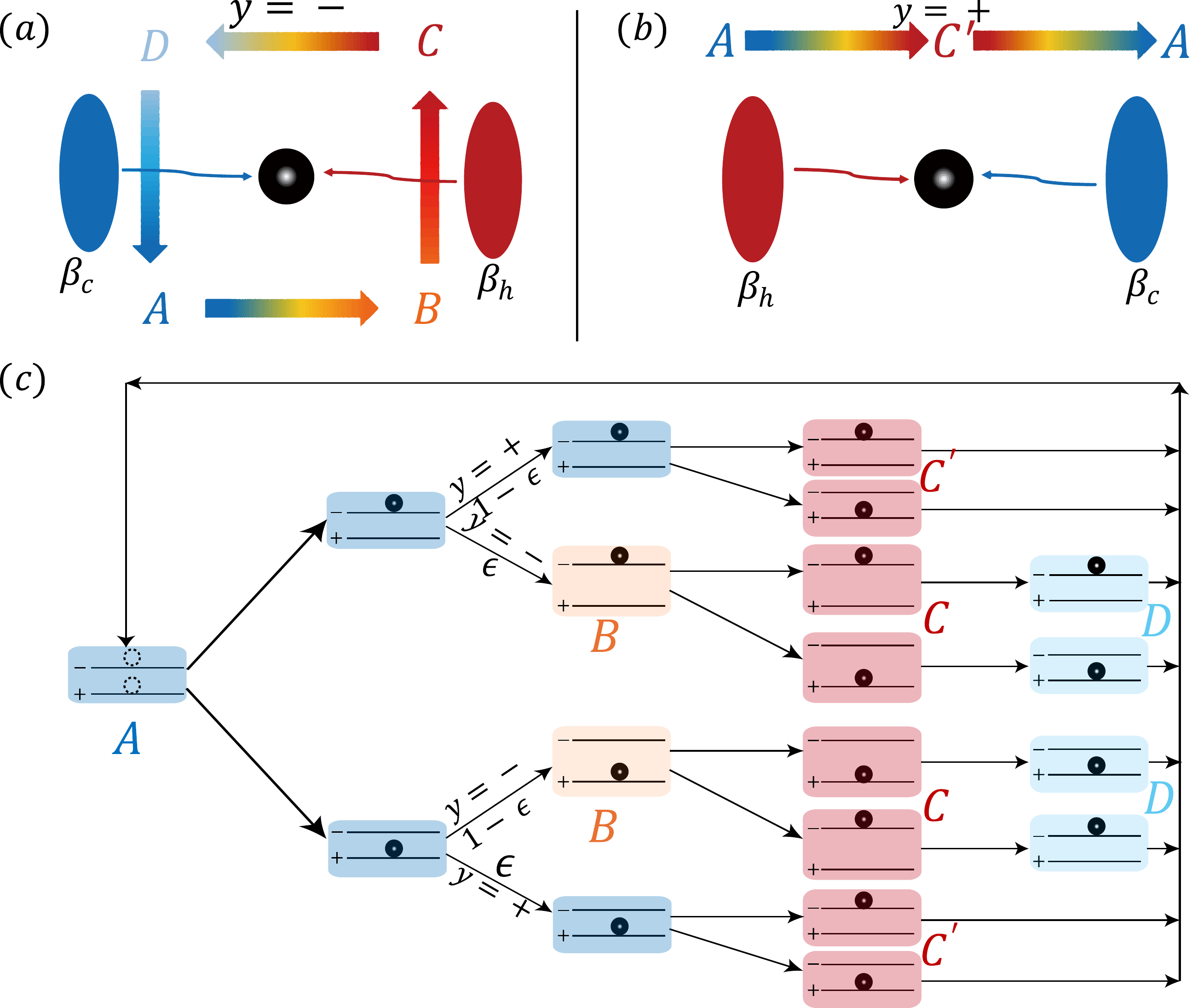}
    \caption{Model of the Otto information engine (OIE). 
    (a) The standard Otto cycle comprising adiabatic compression ($A \to B$), isochoric heating ($B \to C$), adiabatic expansion ($C \to D$), and isochoric cooling ($D \to A$). This sequence also corresponds to the strokes executed in the case of $y=-$ in the OIE. 
    (b) The strokes of the OIE for  $y=+$: isochoric heating ($A \to C^\prime$) followed by isochoric cooling ($C^\prime \to A$). 
    (c) Stochastic trajectories of the OIE in the quasi-static mechanism.}
    \label{model}
\end{figure}

The standard Otto engine (SOE) alternates between two adiabatic strokes ($A \to B$ and $C \to D$) and two isochoric strokes ($B \to C$ and $D \to A$), as illustrated in Fig.~\ref{model}(a).  Here, the working substance is a spin-$1/2$ system and is governed by the Hamiltonian
$H = \frac{\omega}{2}\sigma_z$ ($\hbar=1$), where $\omega$ denotes the external field  frequency and $\sigma_z$ is the Pauli matrix. At equilibrium with a reservoir at inverse temperature $\beta$, the system is described by the Gibbs state $\rho = e^{-\beta H}/\mathrm{Tr}[e^{-\beta H}]$. 
Consequently, the occupation probabilities for the ground state $|-\rangle$ and the excited state $|+\rangle$ can be determined $  p_- = \langle - |\rho| - \rangle = 1/[1+\exp(-\beta\omega)], 
    p_+ = \langle + |\rho| + \rangle = \exp(-\beta\omega)/[1+\exp(-\beta\omega)]$. 
The detail thermodynamic cycle of the quasi-static SOE is  as follows:

First, the system undergoes an adiabatic compression process $A\to B$. During this stroke, the system is isolated from the reservoirs, and its energy level gap is slowly modulated from $\omega_c$ to $\omega_h$. The stochastic work done on the system during this process is denoted as $w_{ch,quai}^{Otto}$. Next, keeping the energy level gap fixed, the system is coupled to a hot reservoir at inverse temperature $\beta_h$ until it reaches a thermal Gibbs state at point $C$. The stochastic heat absorbed by the system from the hot reservoir during this isochoric process is $q_{quai}^{Otto}$. Subsequently, the system is isolated from the reservoirs again, and the energy level gap is slowly decreased back to its initial value at point $D$. The work done by the system on the external environment during this adiabatic expansion stroke is $w_{hc,quai}^{Otto}$. Finally, with the energy level gap fixed at $\omega_c$, the system is coupled to a cold reservoir at inverse temperature $\beta_c$ to return to its initial thermal Gibbs state, thereby completing the cycle.

After the completing one cycle, the total stochastic work produced by the system is $w^{Otto}_{quai} = w_{ch,quai}^{Otto} + w_{hc,quai}^{Otto}$.
During the adiabatic process, the state of the system remains unchanged. After completing one cycle, the distributions of work $w^{Otto}_{quai}$ and heat $q_{quai}^{Otto}$ of the system are as follows (see Appendix \ref{jklg}):

\begin{eqnarray}
    p(w^{Otto}_{quai})
    &=& p_{+}^A p_{-}^C \delta[w^{Otto}_{quai} - (\omega_c - \omega_h)] \nonumber \\
   &+&p_{-}^A p_{+}^C \delta[w^{Otto}_{quai} - (\omega_h - \omega_c)] \nonumber \\
    & +& (p_{+}^A p_{+}^C + p_{-}^A p_{-}^C) \delta(w^{Otto}_{quai}).\label{fghj}\\
    p(q_{quai}^{Otto}) 
    &=& p_{+}^A p_{-}^C \delta(q_{quai}^{Otto} + \omega_h) + 
    p_{-}^A p_{+}^C \delta(q_{quai}^{Otto}- \omega_h) \nonumber \\
    &+& (p_{+}^A p_{+}^C + p_{-}^A p_{-}^C) \delta(q_{quai}^{Otto}).\label{ghj}
\end{eqnarray}
Here, $p_{+}^A p_{-}^C$, $p_{-}^A p_{+}^C$,  $p_{+}^A p_{+}^C$, and  $p_{-}^A p_{-}^C$ represent the probabilities of the SOE running along trajectories $(x_A=+,x_C=-)$, $(x_A=-,x_C=+)$, $(x_A=+,x_C=+)$, and $(x_A=-,x_C=-)$, respectively. The term $x_A$ and $x_C$ denote the system state at points $A$ and $C$, respectively.

Based on Eqs. (\ref{fghj}) and (\ref{ghj}),  the average work output $\langle w^{Otto}_{quai} \rangle$, the work fluctuations $\delta w^{Otto^2}_{quai}= \langle (w^{Otto}_{quai})^2 \rangle-
    \langle w^{Otto}_{quai}\rangle ^2 $, and average  heat absorbed  $\langle q_{quai}^{Otto} \rangle$ from the hot reservoir  can be determined :
\begin{eqnarray}
    \langle w^{Otto}_{quai} \rangle&=&
    (\omega_h - \omega_c)(p_{-}^A p_{+}^C - p_{+}^A p_{-}^C),\label{jl}\\
    \delta w^{Otto^2}_{quai}
    &=& (\omega_h - \omega_c)^2 [(p_{-}^A p_{+}^C + p_{+}^A p_{-}^C)\nonumber \\
    &-& (p_{-}^A p_{+}^C - p_{+}^A p_{-}^C)^2],\label{plg}\\
    \langle q_{quai}^{Otto} \rangle 
    &=&\omega_h (p_{-}^A p_{+}^C - p_{+}^A p_{-}^C).
\end{eqnarray}

Finally, the efficiency  defined as the work divided by the heat of the SOE is 
\begin{equation}
    \eta^{Otto}_{quai} = \frac{\langle w^{Otto}_{quai}\rangle}{\langle q_{quai}^{Otto} \rangle}
     = 1 - \frac{\omega_c}{\omega_h}.
\end{equation}

\section{The Otto information engine}

From Eq. (\ref{fghj}), we can find that work extraction is highly non-uniform along different  trajectories. While the
trajectory  $(x_A=-, x_C=+)$ yields a positive work output of $\omega_h-\omega_c$,  the trajectory $(x_A=+,x_C=-)$ results in a work deficit of $\omega_c-\omega_h$.  Other trajectories $x_A=x_C$  contribute zero net work.
To optimize performance, an effective method is to filter out the detrimental trajectory $(x_A=+,x_C=-)$. To achieve this, we introduced a Maxwell demon into the SOE, called the Otto information engine (OIE). The detailed dynamic cycle of the OIE is as follows.

At point $A$, a Maxwell’s demon is introduced to  measure the system state $x_A$.  The measurement outcome  $y$ has an associated error probability $p(y \neq x_A|x_A) = \epsilon$, accompanying  measurement accuracy being $p(y =x_A|x_A) = 1-\epsilon$. 
After the measurement, the distribution of the demon is  $p(y=+)=p(y=+|x_A=-)p_-^A+p(y=+|x_A=+)p_+^A= \epsilon p_-^A+(1-\epsilon)p_+^A$ and $p(y=-)=p(y=-|x_A=-)p_-^A+p(y=-|x_A=+)p_+^A= (1-\epsilon) p_-^A+\epsilon p_+^A$.
The information obtained by the demon from the system  is  
\begin{eqnarray}\label{agtpl}
    I&=&\sum_{x_A,y}p(y|x_A)p(x_A)\ln \frac{p(y|x_A)}{p(y)}\nonumber\\
    &=&S_Y-S_\epsilon,
\end{eqnarray}
where $S_Y=-p(y=+)\ln p(y=+)-p(y=-)\ln p(y=-)$ represents the demon's entropy after the measurement, and $S_\epsilon=-\epsilon\ln\epsilon-(1-\epsilon)\ln(1-\epsilon)$ is the entropy associated with the measurement error.

Based on the outcome $y$, the demon then implements the following control protocols:  If $y=-$, the demon directs the system to undergo the standard Otto cycle operation, as shown in Fig. \ref{model}(a).
    If $y=+$, the demon bypasses the adiabatic work strokes, directly coupling the system first to the hot reservoir and subsequently to the cold reservoir, as shown in Fig. \ref{model}(b). At point $C^{'}$,  the probability of the system being in  state $x_{C^{'}}$ is $p_{x_{C^{'}}}^{C^{'}}$.

\subsection{The performance of the Otto information engine}

 The stochastic trajectories of the  OIE  are plotted in Fig. \ref{model}(c).  
Under the feedback protocol, the system undergoes the work-producing strokes only when the measurement outcome is $y=-$. 
Consequently, work extraction is restricted to the  trajectories $(y=-, x_A=-, x_C=+)$ and $(y=-, x_A=+, x_C=-)$, yielding positive work $\omega_h-\omega_c$ and negative work $\omega_c-\omega_h$, respectively.
All other trajectories, either bypassed $y=+$ or  $x_A = x_C$ under $y=-$, produce zero net work.
Given the joint probabilities $p(y,x_A,x_C)=p(y|x_A)p(x_A)p(x_C)$,
the probability distribution of the work output per cycle  is
\begin{align}\label{asdfg} 
    p(w_{{quai}}^{{OIE}}) &= \delta[w_{{quai}}^{{OIE}}-(\omega_h-\omega_c)] p(y=-|x_A=-) p_{-}^A p_{+}^C \nonumber \\
    &  + \delta[w_{{quai}}^{{OIE}}+(\omega_h-\omega_c)] p(y=-|x_A=+) p_{+}^A p_{-}^C \nonumber \\
    &  + \delta(w_{{quai}}^{{OIE}})[p(y=-|x_A=-) p_{-}^A p_{-}^C \nonumber \\
    &  + p(y=-|x_A=+) p_{+}^A p_{+}^C + p(y=+)].
\end{align}
Following this work distribution, the average work  output can be obtained  
\begin{align}\label{wOIE}
    \langle w^{{OIE}}_{{quai}} \rangle 
    &=
    (\omega_h-\omega_c)[(1-\epsilon) p_{-}^A p_{+}^C - \epsilon p_{+}^A p_{-}^C ] \nonumber \\
    &= \langle w_{{quai}}^{Otto} \rangle + (\omega_h-\omega_c)[ p_{+}^A p_{-}^C - \epsilon(p_{+}^A p_{-}^C + p_{-}^A p_{+}^C)].
\end{align}
When the measurement of the demon is free error $\epsilon=0$, the trajectory  $(y=-, x_A=+, x_C=-)$ producing negative work will not exist, where the work $\langle w^{{OIE}}_{{quai}} \rangle $ achieves the maximum value $\langle w^{{OIE}}_{{quai}} \rangle= (\omega_h-\omega_c)p_{-}^A p_{+}^C$. Then,
due to $p_{+}^A p_{-}^C + p_{-}^A p_{+}^C>0$, $\langle w^{{OIE}}_{{quai}} \rangle$  is the monotonic decreasing function of $\epsilon$ as widely believed \cite{TK22,GP20,Sag14}. 
Furthermore, Eq. ~(\ref{wOIE}) indicates that the OIE maintains work advantage over the SOE as long as the measurement error remains below relation
\begin{equation}\label{iohgf}
    \epsilon < \frac{p_{+}^A p_{-}^C}{p_{+}^A p_{-}^C + p_{-}^A p_{+}^C}.
\end{equation}
Beyond this limit, the detriment introduced by the measurement error outweighs the benefit of trajectory filtering.

Using Eq. (\ref{asdfg}) again, the corresponding work fluctuations $\delta w^{{OIE}^2}_{{quai}}=\langle(w^{{OIE}}_{{quai}})^2\rangle-\langle w^{{OIE}}_{{quai}}\rangle^2$  can be derived 
\begin{align}
    \delta w^{{OIE}^2}_{{quai}}
    &= (\omega_h-\omega_c)^2 \big\{ [ (1-\epsilon) p_{-}^A p_{+}^C + \epsilon p_{+}^A p_{-}^C ] \nonumber \\
    &- [ (1-\epsilon) p_{-}^A p_{+}^C - \epsilon p_{+}^A p_{-}^C]^2 \big\}. \label{fluct}
\end{align}
Comparing the work distributions Eqs. (\ref{fghj}) and (\ref{asdfg}), the OIE exhibits a reduced probabilities of work  $\omega_h-\omega_c$ and  $\omega_c-\omega_h$ while elevating the zero-work peak. Thus,
by concentrating the probability mass near the origin, the OIE has smaller work fluctuation than the SOE (see Appendix \ref{M2} for mathematical proof).

The heat exchange with the reservoirs is also trajectory-dependent. For paths where $y=-$, the heat absorbed from the hot reservoir along the trajectories $(x_A=-, x_C=+)$, $(x_A=+, x_C=-)$, $(x_A=-, x_C=-)$, and $(x_A=+, x_C=+)$ correspond to the SOE values $\omega_h$, $-\omega_h$, $0$, and $0$, respectively. 
For the bypass trajectories ($y=+$), the heat exchange along the trajectories  $(x_A=-, x_{C^{'}} =+)$, $(x_A=+, x_{C^{'}}=-)$, $(x_A=-, x_{C^{'}}=-)$, and $(x_A=+, x_{C^{'}}=+)$  are $\omega_c$, $-\omega_c$, $0$, and $0$, respectively.
 Summing over all stochastic paths, the probability distribution for the absorbed heat is given by:
\begin{align}
    p(q^{OIE}_{{quai}}) &= \delta(q^{{OIE}}_{{quai}}-\omega_h) p(y=-|x_A=-) p_{-}^A p_{+}^C \nonumber \\
    & + \delta(q^{{OIE}}_{{quai}}+\omega_h) p(y=-|x_A=+) p_{+}^A p_{-}^C \nonumber \\
    & + \delta(q^{{OIE}}_{{quai}}-\omega_c) p(y=+|x_A=-) p_{-}^A p_{+}^{C'} \nonumber \\
    & + \delta(q^{{OIE}}_{{quai}}+\omega_c) p(y=+|x_A=+) p_{+}^A p_{-}^{C'} \nonumber \\
    &+ \delta(q^{{OIE}}_{{quai}})[ p(y=-|x_A=-) p_{-}^A p_{-}^C \nonumber \\
    &+p(y=+|x_A=+) p_{+}^A p_{+}^C \nonumber \\
    &+p(y=-|x_A=-) p_{-}^A p_{-}^{C'} \nonumber \\
    &+ p(y=+|x_A=+) p_{+}^A p_{+}^{C'}].
\end{align}
By this distribution, the average heat exchange with the hot reservoir is
\begin{align}\label{gnm}
    \langle q_{{quai}}^{{OIE}} \rangle &= \int q_{{quai}}^{{OIE}}p(q_{{quai}}^{{OIE}})dq_{{quai}}^{{OIE}}\nonumber \\
    &=\omega_h [ (1-\epsilon) p_{-}^A p_{+}^C - \epsilon p_{+}^A p_{-}^C ] \nonumber \\
    & + \omega_c [ \epsilon p_{-}^A p_{+}^{C'} - (1-\epsilon) p_{+}^A p_{-}^{C'}].
\end{align}
Combining Eqs. (\ref{wOIE}) and (\ref{gnm}), the  efficiency $\eta_{{quai}}^{{OIE}} = \langle w^{{OIE}}_{{quai}} \rangle/\langle q^{{OIE}}_{{quai}} \rangle$ of the OIE reads 
\begin{align}\label{cvb}
    \eta_{{quai}}^{{OIE}} 
    = \frac{\eta^{Otto}_{quai}}{1 + \frac{\omega_c}{\omega_h} \frac{\epsilon p_{-}^A p_{+}^{C'} - (1-\epsilon) p_{+}^A p_{-}^{C'}}{(1-\epsilon) p_{-}^A p_{+}^C - \epsilon p_{+}^A p_{-}^C}}.
\end{align} 
Here, $ \eta^{{OIE}}_{{quai}}$  only considers the heat exchange between the system and the thermal reservoir, without taking into account the cost of the demon, therefore, it is a local efficiency.
According to Eq. (\ref{cvb}), the condition for $\eta_{quai}^{OIE}>\eta^{Otto}_{quai}$ is 
\begin{align}\label{ilk}
    \epsilon< \frac{p_{+}^A p_{-}^{C'}}{p_{-}^A p_{+}^{C'}+p_{+}^A p_{-}^{C'}},
\end{align}
where is the condition for the system injects heat into the hot reservoir when $y=+$.

Due to  $\beta_h\omega_c<\beta_h\omega_h$, we have the relation $p_+^C<p_+^{C^{'}}<p_-^{C^{'}}<p_-^C$, indicating  the right side of Eq. (\ref{ilk}) is smaller than the right side of Eq. (\ref{iohgf}). Consequently, in the regime of Eq. (\ref{ilk}), we have 
\begin{equation}\label{qlk}
    \langle w^{{OIE}}_{{quai}} \rangle > \langle w^{{Otto}}_{{quai}} \rangle,
     \quad \delta w_{{quai}}^{{OIE}^2} < \delta w_{{quai}}^{{Otto}^2}, 
     \quad \eta^{{OIE}}_{{quai}} > \eta^{{Otto}}_{{quai}}.
\end{equation}
where the OIE  achieves a complete superiority over the SOE.

\subsection{The OIE efficiency with the demon' cost}

To ensure the  stable operation of the demon, it’s memory must be erased.  
According to the Landauer limit \cite{Vedral09,Landauer61,Sagawa2009}, the energy required to erase the demon memories using the cold reservoir is $I/\beta_c$.
We therefore focus on the efficiency $\eta_{{quai}}^d$, which accounts for the erasure cost using the cold reservoir
\begin{equation}\label{effdd}
    \eta_{{quai}}^d = \frac{\langle w^{{OIE}}_{{quai}} \rangle}{\langle q^{{OIE}}_{{quai}} \rangle + I/\beta_c}
    = \frac{\eta^{{Otto}}_{{quai}}}{1 + \langle q_-\rangle^{-1}(\langle q_+\rangle + I/\beta_c)}.
\end{equation}
Here, the system and the demon are regarded as a whole, and the output work of the system is the output work of this whole. The energy input to this whole is the heat absorbed by the system from the hot reservoir and the energy required to erase the demon's memory. Thus, $\eta_{{quai}}^d$ is the global efficiency. 
The terms, $\langle q_-\rangle=\omega_h [ (1-\epsilon) p_{-}^A p_{+}^C - \epsilon p_{+}^A p_{-}^C ]$  and $\langle q_+\rangle=\omega_c [ \epsilon p_{-}^A p_{+}^{C'} - (1-\epsilon) p_{+}^A p_{-}^{C'}]$ are the heat absorbed from  the hot reservoir for $y=-$ and $y=+$, respectively.

\begin{figure}[h]
    \centering
    \includegraphics[width=0.9\linewidth]{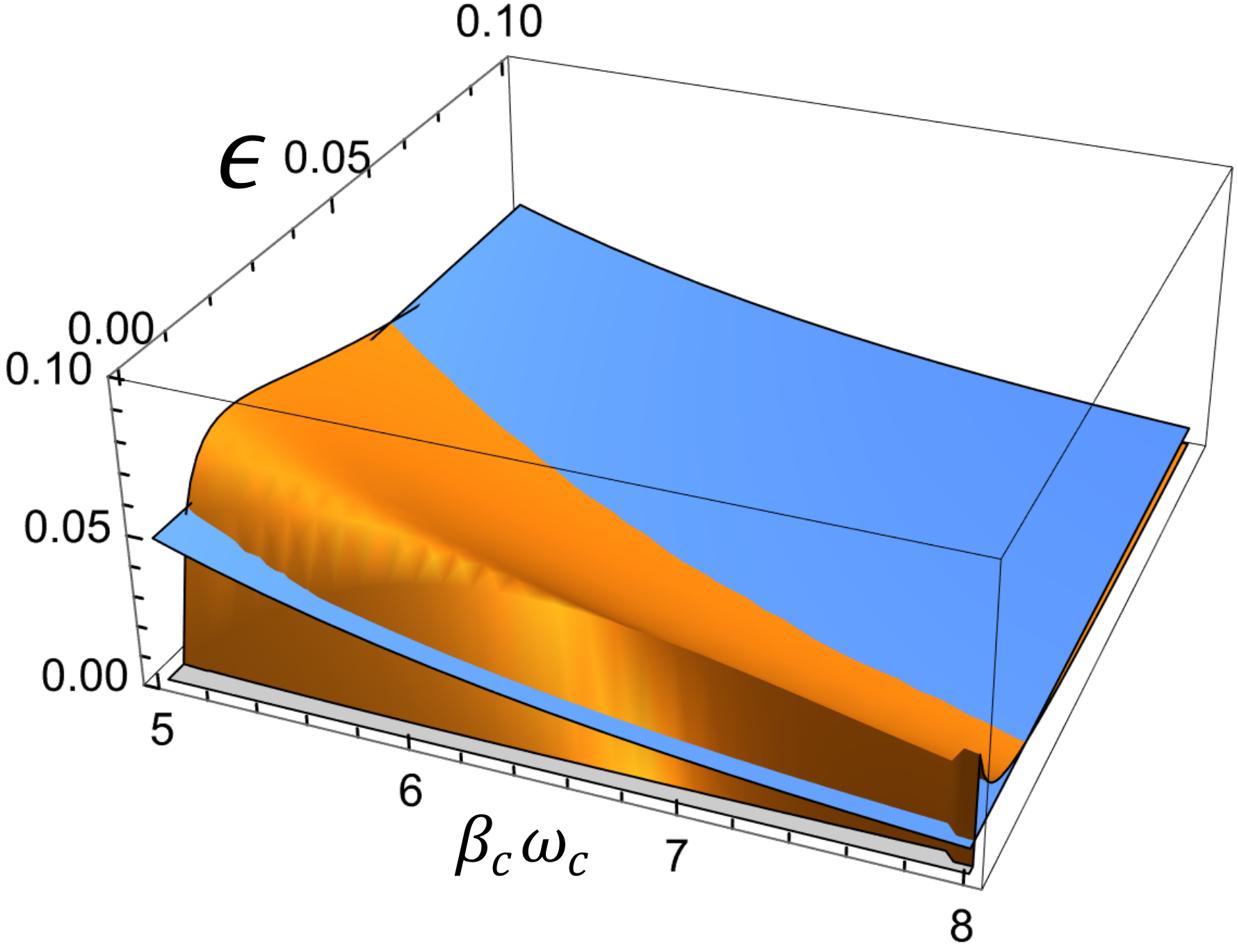}
    \caption{$G$ and $p_{+}^{C'}$ as functions of $\beta_c\omega_c$ and $\epsilon$ for  $\eta_C=0.4$. The yellow surface represents the  function $G$, and the blue surface represents the probability $p_{+}^{C'}$. The region where the blue surface lies below the yellow surface ($p_{+}^{C'} < G$) indicates the parameter space where $\eta_{{quai}}^d>\eta^{{Otto}}_{quai}$.}
    \label{prove1}
\end{figure}

Due to $\langle w_{quai}^{OIE}\rangle>0$ is required, we have $\langle q_-\rangle>0$.
Thus, the condition for the efficiency $\eta_{{quai}}^d$ exceeds the SOC efficiency $\eta^{{Otto}}_{quai}$ is $\beta_c\langle q_+\rangle + I < 0$. 
Substituting the expression of $\langle q_+\rangle$, this condition becomes
    $\beta_c\omega_c[\epsilon p_{-}^A p_{+}^{C'} - (1-\epsilon) p_{+}^A p_{-}^{C'}] + I < 0$, indicating $p_{+}^{C'} < G$ where   $G$ is defined as
\begin{equation}
    G = \frac{\beta_c\omega_c p_{+}^A (1 - \epsilon) - I}{\beta_c\omega_c (p_{-}^A \epsilon + p_{+}^A - \epsilon p_{+}^A)}.
\end{equation}
Clearly, $G$ is a function of the measurement error $\epsilon$ and the parameter $\beta_c\omega_c$. 
Furthermore,  the probability $p_{+}^{C'}$ is the function of the parameter $\beta_h\omega_c$. Due to $\beta_h\omega_c=(1-\eta_C)\beta_c\omega_c$, when the Carnot efficiency $\eta_C=1-\beta_h/\beta_c$
is fixed, $p_{+}^{C'}$ depends on $\beta_c\omega_c$. 
In Fig.~\ref{prove1}, we have plotted $G$ and $p_{+}^{C'}$ as functions of $\beta_c\omega_c$ and $\epsilon$ for a fixed $\eta_C$.
As illustrated in Fig.~\ref{prove1}, there exists a parameter regime where $p_{+}^{C'}$ is  less than $G$. 
This confirms that, even when the demon' cost is included, the efficiency of the OIE can  surpass that of the SOE. This further demonstrates the benefits (comprehensively improving performance) of introducing Maxwell's demon into traditional heat engines.

Furthermore, in Eq. (\ref{effdd}),
the condition for  efficiency $\eta_{quai}^d$ to exceed the Carnot efficiency is $\beta_c \langle w^{{OIE}}_{quai}\rangle + (\beta_h-\beta_c)\langle q^{{OIE}}_{quai}\rangle > \eta_C I$. 
However, according to the Sagawa-Ueda theorem \cite{Sagawa12},  we have  $\beta_c \langle w^{{OIE}}_{quai}\rangle + (\beta_h-\beta_c)\langle q^{{OIE}}_{quai}\rangle < I$. 
Since the Carnot efficiency satisfies $0 < \eta_C < 1$, it is thermodynamically permissible for $\eta_{{quai}}^d$ to surpass the Carnot limit.

{Finally, the local efficiency and the global efficiency satisfies the  relation $\eta_{quai}^{OIE} > \eta_{quai}^d$ due to $I > 0$. Since $\eta_{quai}^d$ can exceed the Carnot efficiency, $\eta_{quai}^{OIE}$ can therefore also exceed the Carnot efficiency. The physical reason is that: according to the Sagawa-Ueda theorem, we have $\eta_{quai}^{OIE} <\eta_{up}^S =\eta_C + I/(\beta_c \langle q_{quai}^{OIE} \rangle)$, whose upper bound can exceed the Carnot limit. Furthermore, tighter upper and lower bounds can be determined based on the information dissipation fluctuation theorem.}
 
\begin{figure*}
    \centering
    \includegraphics[width=1\linewidth]{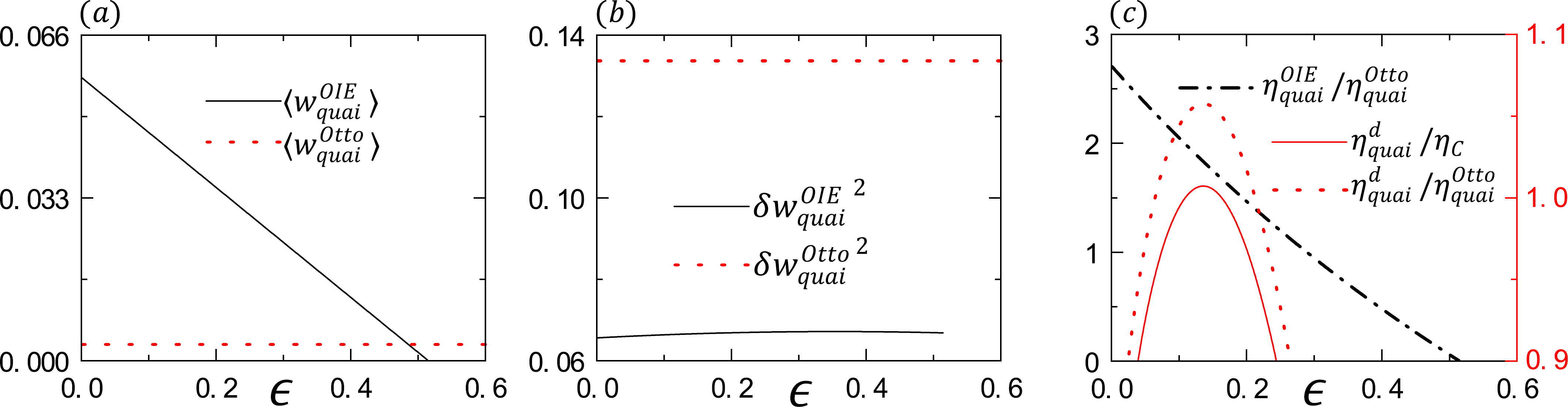}
    \caption{The Performance parameters  of the OIE as  functions of the measurement error $\epsilon$. In (c), the left axis is the scale of $\eta_{quai}^{OIE}/\eta_{quai}^{Otto}$ and the right axis describes $\eta_{quai}^{d}/\eta_{quai}^{Otto}$ and $\eta_{quai}^{d}/\eta_C$. 
    The  parameters are $\omega_h=4.2$, $\omega_c=3$, $\beta_h=0.7$, and $\beta_c=1$.}
    \label{werr}
\end{figure*}
{Since the demon controls the evolution of the system following the measurement, the heat exchange between the system and the hot and the cold reservoirs is affected by the demon. This influence can be quantified as the information flow from the demon to the reservoirs:
$\delta I = \sum_{X_F,y} p(X_F,y) \{ \ln[\frac{p(X_F|x_i,y)}{p(X_B|x_f,y)}] - \ln[\frac{p(X_F|x_i)}{p(X_B|x_f)}]\}$ \cite{Zeng21}. Here, the  forward-time trajectory of the system is defined as $X_F$, with  $x_i$ and $x_f$ being  the initial state and the final state, the  reverse-time trajectory is defined as $X_B$. The sum of this information flow and the information $I$ is referred to  the information  dissipation $\Sigma_I = \delta I + I$ \cite{Zeng21}. Defining the total entropy production of the system, the heat reservoirs, and the demon after completing one cycle as $\Sigma_{tot}=-\beta_h \langle q_{quai}^{OIE} \rangle - \beta_c (\langle w_{quai}^{OIE} \rangle - \langle q_{quai}^{OIE} \rangle) + I$ \cite{Sagawa12}, the efficiency $\eta_{quai}^{OIE}$ can be expressed as $\eta_{quai}^{OIE} = \eta_C + (I - \Sigma_{tot})(\beta_c \langle q_{quai}^{OIE} \rangle)$. Here, we have  employed the cycle condition where the system's average entropy change is zero during a cycle, and the conservation of energy where  the heat absorbed by the system from the cold reservoir is $\langle w_{quai}^{OIE}\rangle-\langle q_{quai}^{OIE}\rangle$. According to the information dissipation fluctuation theorem, we have $\Sigma_{tot} \ge \Sigma_I = \delta I + I \ge 0$, which indicates that $-\beta_h \langle q_{quai}^{OIE} \rangle - \beta_c (\langle w_{quai}^{OIE} \rangle - \langle q_{quai}^{OIE} \rangle) \ge \delta I$ and $I \ge -\delta I$. Consequently, we can obtain the local efficiency  bounds:
\begin{eqnarray}\label{amdk}
    \eta_{low}^Z\le \eta_{quai}^{OIE}
    \le \eta_{up}^Z  \le \eta_{up}^S,
\end{eqnarray}
Here, 
\begin{equation}
    \eta_{low}^Z = \eta_C-\frac{\delta I + \Sigma_{tot}}{\beta_c \langle q_{quai}^{OIE} \rangle},
\end{equation}
 and 
 \begin{equation}
     \eta_{up}^Z = \eta_C - \frac{\delta I}{\beta_c \langle q_{quai}^{OIE} \rangle}
 \end{equation}
  represent the lower   and upper bounds obtained by the information dissipation, respectively. From this inequality, it can be seen that by reducing the total entropy generation $\Sigma_{tot}$, not only can the lower bound  be increased, but also the gap $\eta_{up}^Z-\eta_{low}^Z=\Sigma_{tot}/(\beta_c\langle q_{quai}^{OIE}\rangle)$ between the upper and lower bounds can be reduced. Furthermore, based on the relation $\eta_{up}^Z<\eta_{up}^{S}$, the limiting factor for the local efficiency is the influence of the demons on the heat reservoir, rather than the information that the demons obtain from the system. Finally, if the local efficiency falls within this range,  it does not violate the laws of thermodynamics.}

\subsection{Numerical Analysis}

In Fig.~\ref{werr}(a), we plot the work output of  the OIE and the SOE as functions of the measurement error $\epsilon$. 
Since the SOE operates independently of the measurement outcomes, its work output remains constant. 
However, the OIE work output decreases monotonically with increasing  $\epsilon$   as predicted by Eq.~(\ref{wOIE}). This trend highlights the role of the demon. At $\epsilon=0$, the feedback protocol perfectly filters out trajectories that  yield a work deficit. As $\epsilon$ increases,  the probability of executing these detrimental trajectories rises, suppressing the net work extraction. 
Strikingly, in the error-free limit, the OIE yields approximately an 18-fold increase in work output compared to its standard counterpart, demonstrating the high effectiveness of the information-based trajectory rectification.

\begin{figure}
    \centering
\includegraphics[width=0.8\linewidth]{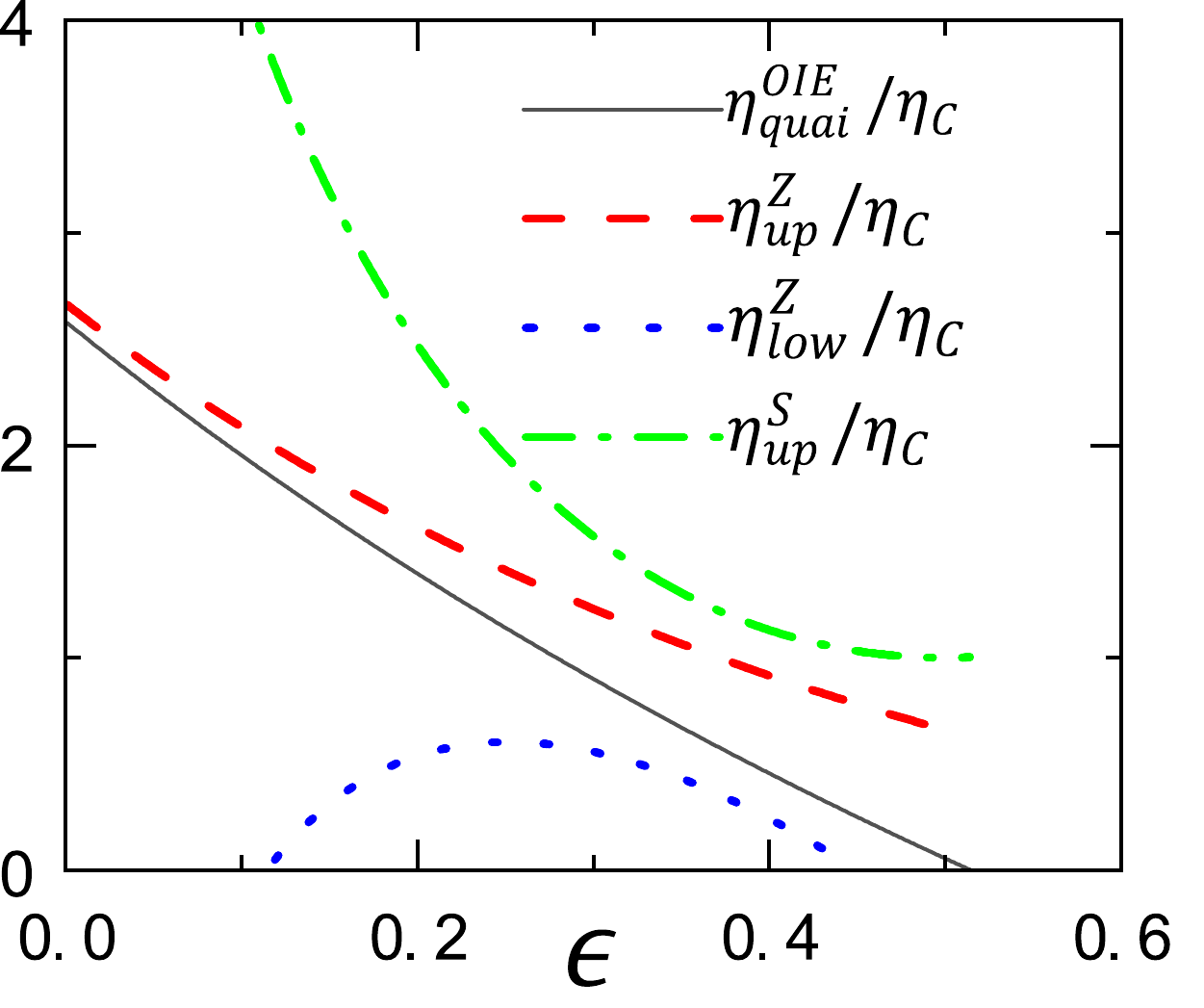}
    \caption{{The local efficiency bounds  as  functions of the measurement error. The paremeters are the same as Fig. \ref{werr}.}}
    \label{ghkler}
\end{figure}

Fig.~\ref{werr}(b) demonstrates that the OIE significantly suppresses work fluctuations, ensuring superior operational stability compared to the SOE, as predicted by Eq. (\ref{fluct}).
\begin{figure*}
    \centering
    \includegraphics[width=1\linewidth]{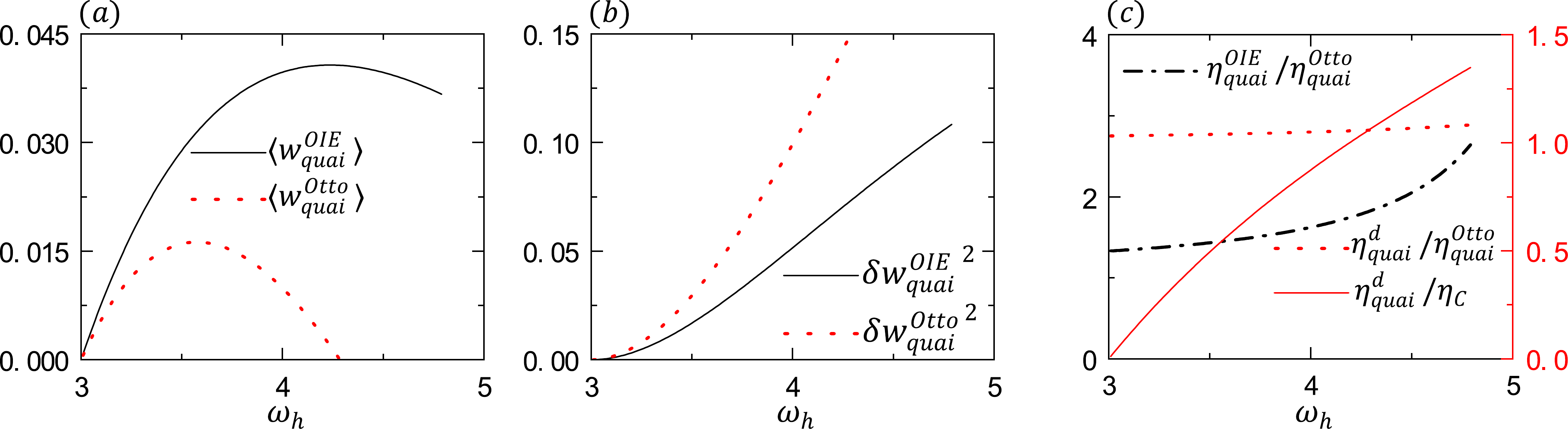}
    \caption{{The OIE performance parameters  as functions of the frequency $\omega_h$, with a fixed measurement error $\epsilon=0.15$. In (c), the left axis is the scale of $\eta_{quai}^{OIE}/\eta_{quai}^{Otto}$. The right axis describes $\eta_{quai}^{d}/\eta_{quai}^{Otto}$ and $\eta_{quai}^{d}/\eta_C$.
    Other parameters are the same as  Fig.~\ref{werr}.}}
    \label{wwh}
\end{figure*}
In the left axis of Fig. \ref{werr}(c), the OIE efficiency significantly exceeds that of the SOE. This enhancement implies that while the demon increases the heat absorbed by the system Eq.
(\ref{gnm}), the gain in work output is more substantial. The right axis of Fig.~\ref{werr}(c) represents the ratios of the OIE efficiency including erasure cost to both the SOC  and Carnot  efficiencies as functions of $\epsilon$. Both ratios exhibit a non-monotonic trend, first increasing and then decreasing with increasing $\epsilon$. This behavior stems from a competition between information gain and work extraction. In the regime $0 < \epsilon < 0.5$, as  $\epsilon$ increases the demon acquires less information, thereby reducing the energetic cost of memory erasure. Initially, this reduction in erasure cost dominates, boosting the efficiency. However, for higher error, the drastic decline in work output due to incorrect feedback takes over, eventually suppressing the efficiency. Remarkably, even  accounting for the demon's cost, the efficiency of the OIE can still exceed the SOC efficiency, and even surpass the Carnot efficiency.
But latter does not violate the second law of thermodynamics. This is because the Carnot efficiency applies to the heat exchange between the system and two heat reservoirs. For example, if we use two heat reservoirs to erase  the  demon' memory and include the cost of the demon in the efficiency, the efficiency will not exceed the Carnot efficiency \cite{TT25}. However, in the OIE, although the working system exchanges heat with two heat reservoirs, the demon only exchanges heat with the cold reservoir. Therefore, the efficiency of the OIE containing the demon cost is not limited by the  Carnot efficiency. {As a result, the local efficiency $\eta^{OIE}_{quai}$ can also exceed the Carnot boundary, as shown in Fig. \ref{ghkler}. Furthermore, in this figure, the local efficiency falls between $\eta_{low}^Z$ and $\eta_{up}^Z$, indicating that our results are reasonable. Finally, since $\eta_{up}^Z$ is less than $\eta_{up}^S$, Fig. \ref{ghkler} also verifies the information dissipation fluctuation theorem.}

{Overall, the output work, the  stability, and the efficiency of the OIE are better than those of the corresponding SOE.}

{In Figs.~\ref{wwh}(a)-(c), the OIE  has also higher work output, suppressed fluctuations, and enhanced efficiency compared to the SOE as frequency $\omega_h$ varies. These results underscore again the  thermodynamic advantage of the OIE.}

{To elucidate the work trends in Fig.~\ref{wwh}(a), we consider the simplified expressions for average work Eqs. (\ref{jl}) and (\ref{wOIE}): $\langle w_{quai}^{Otto}\rangle=(\omega_h-\omega_c)(p^C_+-p^A_+)$ and  $\langle w_{quai}^{OIE}\rangle=(\omega_h-\omega_c)[p(y=-)p^C_+-\epsilon p^A_+]$. Both quantities exhibit a non-monotonic dependence on $\omega_h$ governed by a competition between the energy gap and the population difference. In the small $\omega_h$ regime, the linear increase in $\omega_h-\omega_c$  dominates over the slow decline of the excited-state population $p_+^C$ leading to an initial rise in work. However, as $\omega_h$  continues to increase, $p_+^C$ rapidly approaches its lower bound, causing the population difference to vanish. Consequently, the work output for both engines reaches a maximum and subsequently declines.}

{Notably, in Fig.~\ref{wwh}(a)  the OIE significantly extends the operating regime of the SOC: The SOE fails to extract work for $\omega_h>4.3$. However, the OIE remains functional up to  $\omega_h>4.7$ demonstrating its robustness under strong driving fields. Furthermore, Fig.~\ref{wwh}(c) shows that the OIE efficiency consistently exceeds its standard counterpart, even when accounting for the demon's cost. This advantage is rooted in the condition $p_+^{C^{'}}=0.11<G=0.15$. Critically, combining Figs. ~\ref{wwh}(a) and (c), the OIE maintains its superior efficiency even at the  maximum work output.}

\section{Finite time Case}

The isochoric processes of the finite time SOE  (OIE) are the same as that of the quasi-static SOE (OIE), but a different adiabatic process. The adiabatic processes of the SOE (OIE) with finite time will undergo non adiabatic quantum transitions due to the presence of internal friction, with the transition probability defining as  $\xi$ \cite{RJ19,JPS19}. After completing such a cycle, the output work, work fluctuations, and heat absorption of the SOE  are respectively (see Appendix \ref{CCCC})

{\begin{eqnarray}
\langle w_{fin}^{Otto}\rangle
&=&\langle w_{quai}^{Otto}\rangle+\langle
w_{fri}^{Otto}\rangle,\label{waf}
\end{eqnarray}
\begin{eqnarray}
    \delta w_{fin}^{Otto^2}
   &=& \omega_h^2[\frac{1}{2}-\frac{(p_+^A-p_-^A)^2(1-2\xi)^2+(p_+^C-p_-^C)^2}{4}]\nonumber\\
   &+& \omega_c^2[\frac{1}{2}-\frac{(p_+^A-p_-^A)^2+(p_+^C-p_-^C)^2(1-2\xi)^2}{4}]\nonumber\\
&+&\omega_c\omega_h[1-\frac{(p_+^A-p_-^A)^2+(p_+^C-p_-^C)^2}{2}],\label{jkl}\\
\langle q_{fin}^{Otto}\rangle
   &=&\omega_h[p_-^{A} p_+^{C}- p_+^{A} p_-^{C} + (p_+^{C} - p_-^{C})\xi].\label{adc}
\end{eqnarray}
In Eq. (\ref{waf}),  $\langle w_{fri}^{Otto}\rangle=[(p_+^{A}- p_-^{A}) \omega_c 
+ (p_+^{C} - p_-^{C} ) \omega_h] \xi$ represents the work generated due to internal friction, defined as the work difference  between the actual adiabatic stroke and the ideal one \cite{Plastina14}. 
Obviously, this work is a negative work because $p_+^ {A}<p_-^{A}$ and $p_+^{C}<p_-^{C}$. Consequently, the presence of the internal friction inevitably diminishes the work output of the SOE.  Due to the transition probability $\xi$  increases as the driving time decreases \cite{RJ19,JPS19}, faster driving cycles exacerbate frictional losses,  suppressing the net work extraction.}

{Finally, according to  Eqs. (\ref {waf}) and (\ref {adc}), the efficiency of this finite time SOE is
\begin{eqnarray}\label{bnm1}
\eta_{fin}^{Otto}=
\frac{\langle w_{fin}^{Otto}\rangle}{\langle q_{fin}^{Otto}\rangle}
=\eta^{Otto}_{quai}+\frac{\omega_c\xi(p_+^A-p_-^A+p_+^C-p_-^C)}{\langle q_{fin}^{Otto}\rangle}.
\end{eqnarray}
Obviously, quantum internal friction  also reduces the efficiency of SOE due to $p_+^A<p_-^A$ and $p_+^C<p_-^C$.}

{At the same time, the performance parameters of OIE are also derived in Appendix \ref{CCCC}, and their specific expressions are as follows}

\begin{figure*}
\centering
 \begin{overpic}[width=0.8\linewidth]{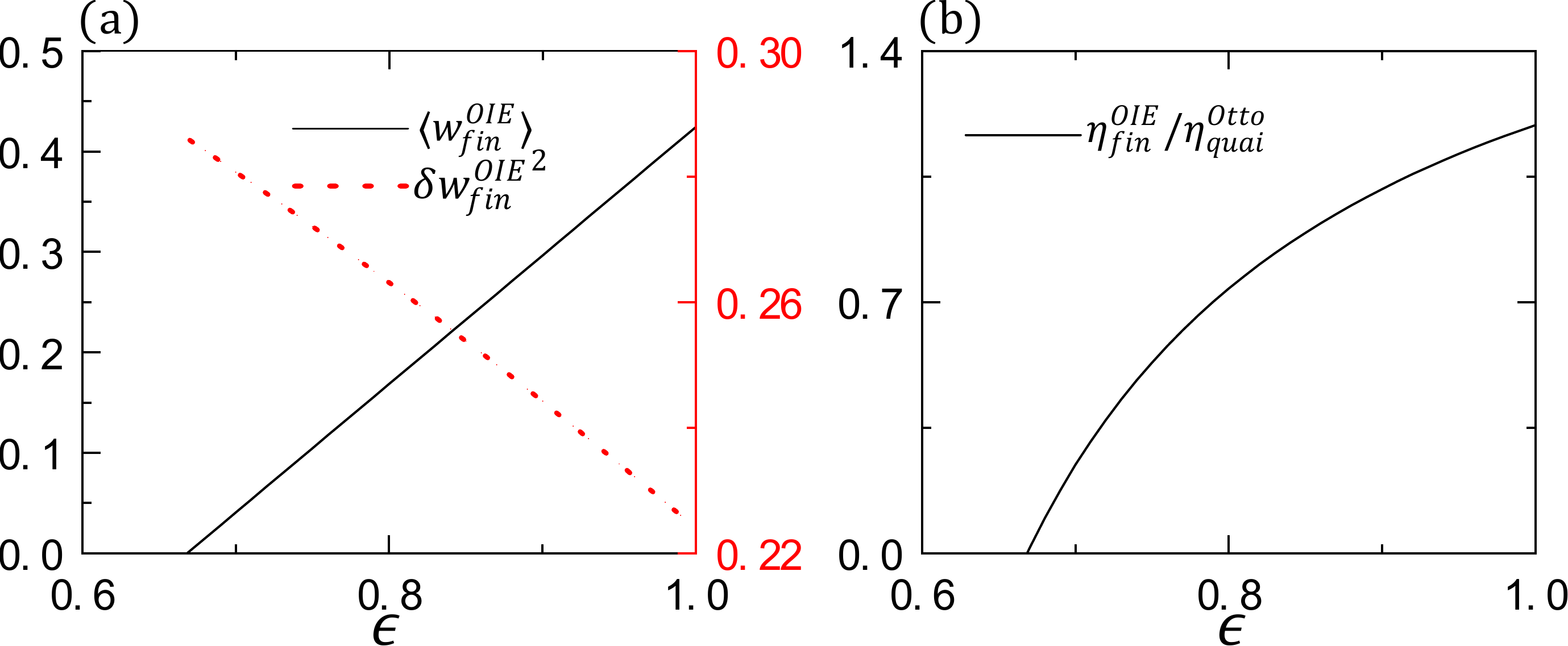}
\end{overpic}
\caption{Performance parameters of the finite time OIE as functions of the  measurement error $\epsilon$ with $\tau=0.1$. In panel (a), the left axis (black) represents output work, and the right axis (red) represents work fluctuations. The parameters are $\beta_h = 0.1$, $\beta_c = 1$, $\omega_h=1,\omega_c=0.5$.}\label{bnmm}
\end{figure*}

\begin{figure*}
\centering
 \begin{overpic}[width=0.8\linewidth]{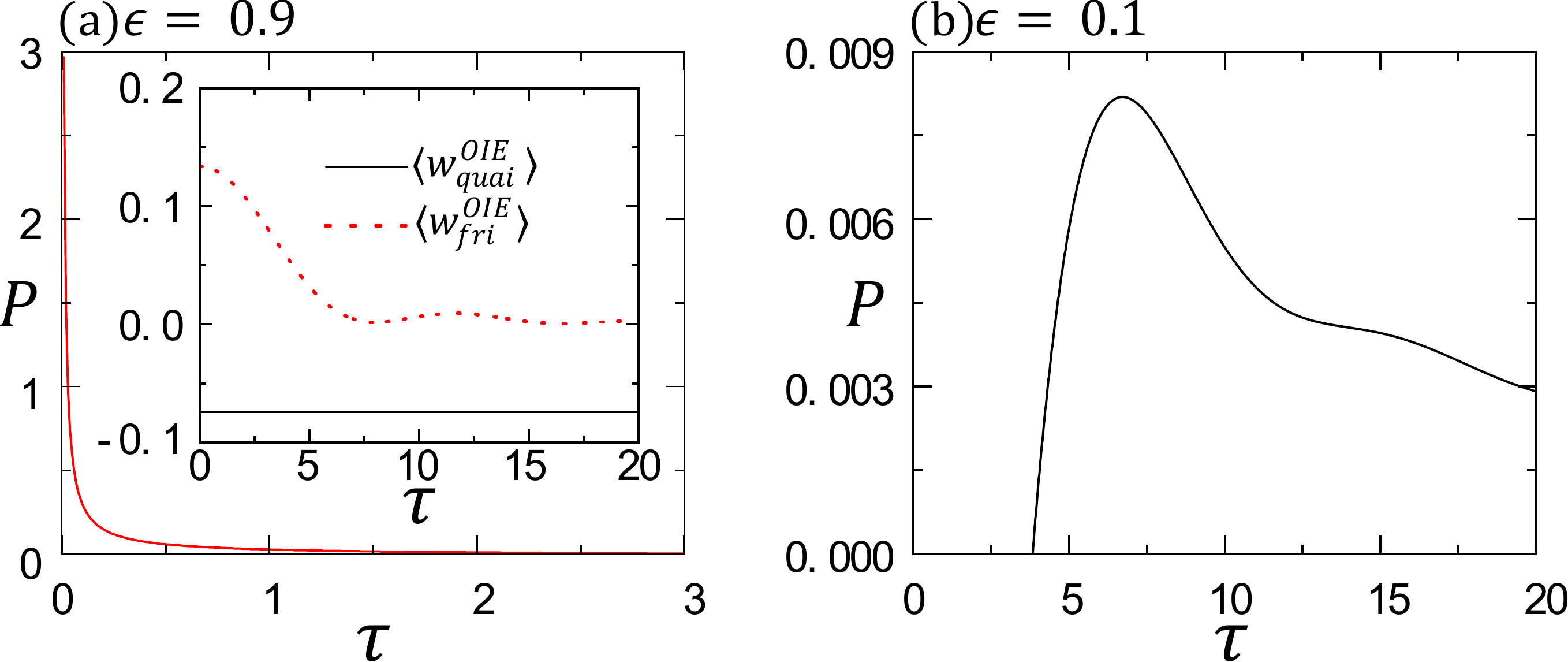}
\end{overpic}
\caption{Power output of the finite time OIE as a function of the driving time $\tau$.
The  parameters are the same as Fig. \ref{bnmm}.
}\label{jnk}
\end{figure*}

\begin{eqnarray}
    \langle w^{OIE}_{fin}\rangle&=&p_-^{A} p_+^{C} (\omega_h-\omega_c) + p_-^{A} [(p_+^{C} -p_-^{C})\omega_c-\omega_h]\xi\nonumber\\
    &+&\{(p_-^{A} p_+^{C}+p_+^{A} p_-^{C}) (\omega_c - \omega_h)\nonumber\\
    &+&[\omega_h+(p_+^{A}-p_-^{A})(p_+^{C} - p_-^{C})\omega_c]\xi\}\epsilon,\label{vbnm}\\
    \delta w^{OIE^2}_{fin}&=& \int p(X_F, y=g)(w_{fin}^{OIE}(X_F,y=g))^2d X_F\nonumber\\
&-& \langle w^{OIE}_{fin}\rangle^2\nonumber\\
   &=& (1-\xi)^2 (\omega_c - \omega_h)^2  [(1-\epsilon) p_-^A p_+^C + \epsilon p_+^A p_-^C]\nonumber \\
& +& \xi^2 (\omega_c + \omega_h)^2 [ \epsilon p_+^A p_+^C + (1-\epsilon) p_-^A p_-^C] \nonumber\\
& +& \xi(1-\xi) (\omega_c^2 + \omega_h^2) [ \epsilon p_+^A + (1-\epsilon) p_-^A ] \nonumber\\
& -& \{ (1-\xi)(\omega_c - \omega_h) [ \epsilon p_+^A p_-^C - (1-\epsilon) p_-^A p_+^C ]\nonumber\\
 &+& \xi(\omega_c + \omega_h) [\epsilon p_+^A p_+^C - (1-\epsilon) p_-^A p_-^C ] \}^2,\label{dwof}\\
\langle q_{fin}^{OIE}\rangle&=&\int p(X_F,y)q_{finite}^{OIE}(X_F,y)dX_Fdy\nonumber\\
&=&\omega_h[(1-\epsilon) p^{A}_- p^{C}_+-\epsilon p^{A}_+ p^{C}_-+(\epsilon-p^{A}_-)\xi]\nonumber\\
&+&\omega_c[p^{A}_- \epsilon p_+^{C^{'}} - p^{A}_+ (1 - \epsilon) p_-^{C^{'}}].\label{jda}
\end{eqnarray}
From Eqs.~(\ref{vbnm}) and (\ref{jda}),  the local efficiency  $\eta_{fin}^{OIE}=\langle 
w^{OIE}_{fin}\rangle/\langle q_{fin}^{OIE}\rangle$ and the global efficiency of containing the demon cost $\eta_{fin}^{d}=\langle w^{OIE}_{fin}\rangle/(\langle q_{fin}^{OIE}\rangle+I/\beta_c)$ of the finite time OIE can be determined. Since the work, fluctuations, and efficiency of both the SOE and the OIE are continuous functions of the transition probability $\xi$, the advantages demonstrated in the quasistatic limit $\xi=0$ are preserved  for small $\xi$. Because the improvement of the   demon on the performance of the finite time Otto engine  is an extension of quasi-static case, we will not discuss it in detail here, but have already included it in Appendix \ref{CCCC}.  In this section, we are concerned with how the quantum inner friction converts the measurement error into a useful resource, as well as how the measurement error converts the quantum inner friction into an available resource

 Eq. (\ref{vbnm}) shows that the coefficient of the  error term scales monotonically with the transition probability  $\xi$.
In the adiabatic limit $\xi\to 0$,  this coefficient takes the value $(p_-^{A} p_+^{C}+p_+^{A} p_-^{C}) (\omega_c - \omega_h)<0$, which returns to Eq. (\ref{wOIE}) where the measurement error reduces the work output. 
Conversely, at maximal transition probability $\xi=0.5$ \cite{RJ19,JPS19}, the coefficient becomes $[\omega_c + (p_+^{A} - p_-^{A}) (p_+^{C} - p_-^{C})\omega_h]/2$ which is positive. This sign change defines a critical threshold $\xi_0=(p_-^{A} p_+^{C}+p_+^{A} p_-^{C})
 (\omega_h - \omega_c)/[\omega_h+(p_+^{A}-p_-^{A})(p_+^{C} - p_-^{C})\omega_c]$, separating two distinct operating regimes. For 
$0<\xi<\xi_0$ the work output decreases with measurement error as expected. However, for $\xi_0<\xi<0.5$ a counter-intuitive reversal occurs: the output work increases alongside the error. In this regime, quantum friction effectively repurposes measurement error, transforming a typically detrimental factor into a constructive resource for work extraction. 

In addition, Eq. (\ref{vbnm}) can be rewritten as 
\begin{eqnarray}\label{wof}
    \langle w^{OIE}_{fin}\rangle
    &=&(\omega_h-\omega_c )[(1-\epsilon) p^{A}_- p^{C}_+-\epsilon p^{A}_+ p^{C}_-]\nonumber\\
    &+&\{p_-^{A} (p_+^{C}  - p_-^{C}) \omega_c - p_-^{A}\omega_h\nonumber\\ 
&+& \epsilon [(p_+^{A}- p_-^{A}) (p_+^{C}  - p_-^{C}) \omega_c + \omega_h]\}\xi\nonumber\\
&=&\langle w_{quai}^{OIE}\rangle+\langle w_{fri}^{OIE}\rangle,
\end{eqnarray}
where $\langle w_{fri}^{OIE}\rangle=\{p_-^{A} (p_+^{C}  - p_-^{C}) \omega_c - p_-^{A}\omega_h 
+ \epsilon [(p_+^{A}- p_-^{A}) (p_+^{C}  - p_-^{C}) \omega_c + \omega_h]\}\xi$  represents the work contribution from the internal friction in the OIE. Crucially, $\langle w_{fri}^{OIE} \rangle $ is an increasing function of the measurement error $\epsilon$, suggesting that imperfect measurement can  mitigate the deleterious effects of the friction.
By setting 
 $\langle w_{fri}^{OIE} \rangle=0 $, we identify a critical error threshold 
 \begin{eqnarray}
\epsilon_f=\frac{p_-^{A}\omega_h - p_-^{A} (p_+^{C} - p_-^{C}) \omega_c}{(p_+^{A} - p_-^{A}) (p_+^{C} - p_-^{C})\omega_c + \omega_h}.
 \end{eqnarray}
 When $\epsilon<\epsilon_f$ the internal friction is harmful to OIE' work output.  However, when $\epsilon>\epsilon_f$, the quantum internal friction transitions from a source of dissipation into a resource for positive work extraction. In this regime, the work generated by the friction is positive and   scales with the transition probability $\xi$ which increases as the driving time decreases. 
 This counterintuitive behavior demonstrates that quantum friction, typically considered a fundamental bottleneck for heat engines, can be harnessed as a valuable thermodynamic resource under the OIE framework. 
 
 One important application of this result is significantly increasing the output power (work/cycle time) of the OIE.
 In the sudden limit $\tau\to 0$, the transition probability $\xi\to 0.5$ yielding a finite and constant work output. Consequently, the power output is primarily limited by the reservoir coupling time. In the high-density photon environments where thermalization is instantaneous, this mechanism enables a dramatic enhancement, or even a formal divergence (ignoring the thermal coupling time), of the OIE power output.

 In Fig.~\ref{bnmm}, the OIE harnesses the  measurement error. Contrary to the quasi-static case where the  error typically degrades performance, the work output and efficiency here increase with the error while the fluctuations decrease as the error grows. This confirms that quantum inner friction can repurpose measurement error, transforming a detrimental  source into a constructive resource that comprehensively enhances the engine's performance.

Then, Fig.~\ref{jnk} shows that the power output  
$P$ as a function of the driving time. In Fig~\ref{jnk}(a), $P$ decreases initially rapidly and then slowly with increasing $\tau$, tracing the decay profile of the quantum inner friction (as shown in the inset). Crucially, the inset reveals that the frictional contribution to work is positive, indicating that the demon actively harnesses the  non-adiabatic transitions to extract work, which is impossible for the  SOE [see Eq.~(\ref{waf})]. Furthermore, while the OIE yields a negative work  in the quasistatic limit for the parameters in Fig.~\ref{jnk}, the quantum inner friction compensates for this negative work, enabling net work extraction.  Conversely, Fig.~\ref{jnk}(b) depicts the regime where the friction acts as a source of dissipation. {In the short-time regime, the power output is primarily governed by the sensitive dependence of internal friction on the driving duration. Initially, $P$  increases with $\tau$ since the rapid reduction of the  friction outweighs the increase in cycle time. However, as $\tau$ continues to grow, the reduction in frictional loss becomes less pronounced where the output work approaches to the quasi-static limit. In this region, the $1/\tau$ scaling in the definition of power becomes the dominant factor, leading to the eventual monotonic decay of $P$ with driving time.}
 Comparing  Figs. ~\ref{jnk}(a) and (b), we find  a dramatic enhancement: the power output in the regime where friction is harnessed  is three orders of magnitude larger than in the regime where friction is detrimental.
Additionally, in Fig. ~\ref{jnk}(a), the output power is divergent due to the neglect of the thermal coupling time.

\section{conclusion}

We have introduced an Otto Information Engine (OIE) that fundamentally transcends the operational limits of the standard Otto engine. Through a selective feedback protocol, the OIE enables a simultaneous enhancement of work output,  stability, and  efficiency.  Crucially, even accounting for the energetic costs of the demon, the OIE efficiency surpasses both the standard Otto  and Carnot efficiencies. {In addition, to demonstrate that our results comply with thermodynamic laws, we obtained the efficiency upper and lower bounds through the information dissipation fluctuation theorem.}

Beyond performance gains, we have identified a novel mechanism where  the  measurement error is transformed from a detrimental  source into a constructive resource which enhances the robustness of the machine.  Furthermore, within the OIE framework, the quantum inner friction, conventionally a damage of work, is repurposed to drive work extraction, bringing  orders of magnitude increase and even  enabling a formal divergence  (ignoring the thermal coupling time) of the power output. These findings challenge the prevailing paradigm that dictates the strict suppression of  the error   and friction  in  engines design.

Finally, with the recent experimental advance in the trapped $^{40}\mathrm{Ca}^+$  technology, including the realization of the  Maxwell’s demons \cite{Yan24}  and  heat engines \cite{Lutz16,Lutz24}, the OIE protocol can be readily implemented. Our work offers new perspectives on the ultimate bounds of stochastic thermodynamics and provides a blueprint for the next generation of precision quantum nanomachines.

\begin{acknowledgments}
 Y. Xiao thanks the support in part by the National Natural Science Foundation of China Grant No. NSFC 12234019, and the Science and Technology Development Plan Project of Jilin Province No. SKL202302029.  
\end{acknowledgments}

\appendix

\section{The detailed thermodynamic cycle of standard Otto heat engine}\label{jklg}
The detailed cycle of the standard Otto heat engine is as follows.

\textit{Adiabatic compression} ($A \to B$): 
Starting from thermal equilibrium with the cold reservoir,  the external frequency is adiabatically and slowly modulated from $\omega_c$ to $\omega_h$. This unitary process preserves the system state. The work generated during this stroke equals the change in the system's internal energy. 
Consequently, the probability distribution of the work performed by the system is given by
\begin{equation}
    p(w_{ch,quai}^{Otto}) = \sum_{x_A} p_{x_A}^A \delta\left[w_{ch,quai}^{Otto}-\frac{x_A(\omega_c - \omega_h)}{2}\right],
\end{equation}
where $p_{x_A}^A$ denotes the probability of the system being in state $x_A$ ($x_A =-, +$) at point $A$.

\textit{Isochoric heating ($B \to C$):} 
During this stroke, the system is coupled to the hot reservoir with the fixed frequency $\omega_h$. 
At the final point $C$, the system reaches thermal equilibrium with the hot reservoir. At this stage, the probability of the system being in state $x_C$ is $p_{x_C}^C$. As no work is performed, the absorbed heat  equals the internal energy change. 
Conditioned on the work $w_{ch,quai}^{Otto}$, the heat distribution  is
\begin{equation}
p(q_{quai}^{Otto}|w_{ch,quai}^{Otto})=\sum_{x_C}p_{x_C}^C \delta\left[q_{quai}^{Otto}-\frac{\omega_h(x_C-x_A)}{2}\right].
\end{equation}
Here, we have used the adiabatic condition  $x_B=x_A$   with $x_B$ being the system state at point $B$.

\textit{Adiabatic expansion ($C \to D$):} 
The frequency is adiabatically returned to  $\omega_c$. 
Similar to the first stroke, the work produced during this stroke  is determined by the energy difference while the system state  remains invariant $x_D=x_C$ with $x_D$ denoting the system state at point $D$. 
The work distribution conditioned on $w_{ch,quai}^{Otto}$ and  $q_{quai}^{Otto}$ reads
\begin{equation}
    p(w_{hc,quai}^{Otto} | q_{quai}^{Otto}, w_{ch,quai}^{Otto}) = \delta\left[w_{hc} - \frac{x_C(\omega_h - \omega_c)}{2}\right].
\end{equation}

\textit{Isochoric cooling ($D \to A$):} 
The cycle is closed by coupling the system to the cold reservoir at frequency $\omega_c$, and eventually returns to the initial thermal distribution.

The total work generated per cycle is $w^{Otto}_{quai} = w_{ch,quai}^{Otto} + w_{hc,quai}^{Otto}$.
 By combining the above processes, the joint probability distribution for the total work and the heat absorbed from the hot reservoir  can be obtained  

 \begin{eqnarray}
    p(w^{Otto}_{quai}, q_{quai}^{Otto}) &=& \sum_{x_A, x_C} p_{x_A}^A p_{x_C}^C\nonumber\\
   &\times&  \delta
    [w^{Otto}_{quai}- \frac{(\omega_h - \omega_c)(x_C-x_A)}{2}]\nonumber\\
   &\times&  \delta[q_{quai}^{Otto} -\frac{\omega_h(x_C - x_A)}{2}].
\end{eqnarray}

The marginal distribution of the work $p(w_{quai}^{Otto})$  is then determined by integrating over the heat $q_{quai}^{Otto}$
\begin{eqnarray}\label{fghj1}
    p(w^{Otto}_{quai}) &=& \int p(w^{Otto}_{quai}, q_{quai}^{Otto}) dq_{quai}^{Otto} \nonumber \\
    &=& p_{+}^A p_{-}^C \delta[w^{Otto}_{quai} - (\omega_c - \omega_h)]\nonumber\\
   &+& 
    p_{-}^A p_{+}^C \delta[w^{Otto}_{quai} - (\omega_h - \omega_c)] + (p_{+}^A p_{+}^C \nonumber\\
   &+&  p_{-}^A p_{-}^C) \delta(w^{Otto}_{quai}).
\end{eqnarray}

Similarly, by integrating the work, the marginal distribution of the heat absorbed by the system from the hot reservoir can be derived
\begin{eqnarray}\label{ghj1}
    p(q_{quai}^{Otto}) &=& \int p(w^{Otto}_{quai}, q_{quai}^{Otto}) dw^{Otto}_{quai} \nonumber \\
    &=& p_{+}^A p_{-}^C \delta(q_{quai}^{Otto} + \omega_h) \nonumber\\
   &+& 
    p_{-}^A p_{+}^C \delta(q_{quai}^{Otto}- \omega_h) \nonumber\\
   &+&  (p_{+}^A p_{+}^C + p_{-}^A p_{-}^C) \delta(q_{quai}^{Otto}).
\end{eqnarray}

\section{The proof of the  OIE being more stable than the SOE under any measurement error}\label{M2}
We now demonstrate that the work fluctuations in the OIE are strictly lower than those of the SOE, $\delta w_{{quai}}^{{OIE}^2} < \delta w_{{quai}}^{{Otto}^2}$, for any measurement error. 
Let $p_p = p_{-}^A p_{+}^C$ and $p_n = p_{+}^A p_{-}^C$ denote the trajectory probabilities of the positive and negative work in the SOE, respectively. 
In the SOE,   $\langle w_{quai}^{Otto} \rangle > 0$ is held, indicating  $p_p > p_n$. Combining this relation and the positive temperature condition, the relations   $0 < p_{+}^A < p_{+}^C < 1/2 < p_{-}^C < p_{-}^A < 1$ can be determined,  resulting $0 < p_n < p_p < 1/2$.

\begin{figure}[h]
    \centering
    \includegraphics[width=1\linewidth]{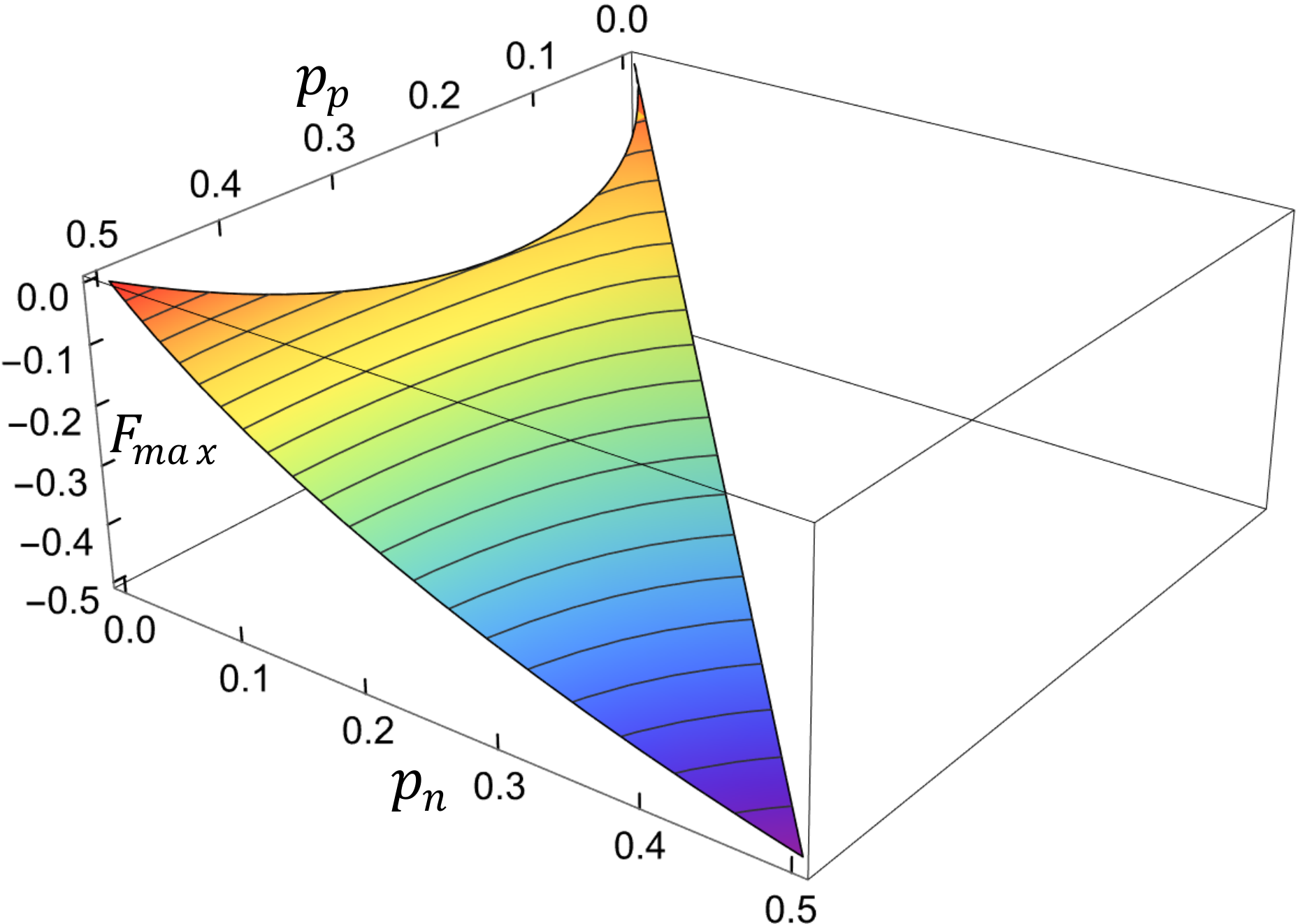}
    \caption{$F_{max}$ as a function of $p_p$ and $p_n$ with   $0<p_n<p_p<1/2$ and $0<\epsilon_s<1$.}
    \label{pr}
\end{figure}

Combing Eqs. (\ref{plg}) and (\ref{fluct}),  the difference of the work fluctuations between  the OIE and the SOE  can be written as  
 $\Delta = (\omega_h-\omega_c)^2 F(\epsilon),$
where 
$F(\epsilon)= [ (1-\epsilon)p_p + \epsilon p_n - ( (1-\epsilon)p_p - \epsilon p_n )^2 ] 
    -[ p_p + p_n - (p_p - p_n)^2]$.
To prove $\Delta<0$, it suffices to show that $F < 0$.
Expanding $F$ a quadratic function of $\epsilon$ can be obtained:
    $F(\epsilon) = -\epsilon^2 (p_p + p_n)^2 + \epsilon (p_n - p_p + 2 p_n p_p + 2 p_p^2)  - p_n - p_p^2 + (p_p - p_n)^2.$ 
At the boundaries $\epsilon=0$ and $\epsilon=1$, the function $F$ remains negative: $F(0)= p_n(p_n -1-2 p_p)<0$ and  $F(1)= p_p(p_p-1 - 2p_n)< 0$.
Furthermore, the symmetry axis of $F(\epsilon)$ lies at $\epsilon_s = (p_n - p_p + 2 p_n p_p + 2 p_p^2)/[2 (p_p + p_n)^2]$.
Due to $F(\epsilon)$ is a downward-opening parabola $-(p_p + p_n)^2<0$,
it increases for $\epsilon < \epsilon_s$ and decreases for $\epsilon > \epsilon_s$.

Based on the above properties, the sign of $F$ can be discussed. If $\epsilon_s < 0$,  $F(\epsilon) < F(0) < 0$ holds for all $\epsilon > 0$. Similarly, if $\epsilon_s > 1$,  then $F(\epsilon) < F(1) < 0$, for all $\epsilon < 1$.
When the symmetry axis lies in  $0 < \epsilon_s < 1$, the function attains its maximum $F_{{max}}$ at $\epsilon = \epsilon_s$:
\begin{equation}
F_{{max}} = \frac{1}{4} - p_p + \frac{(p_n - 1) p_n^3 - p_n p_p + (1 - 2 p_n) p_n p_p^2 + p_p^4}{(p_n + p_p)^2}.
\end{equation}
Numerical evaluation of $F_{\text{max}}$ as a function of  $p_n$ and $p_p$  is plotted in Fig.~\ref{pr}. The maximum value approaches zero only in two asymptotic limits: (i) $p_n \to 0, p_p \to 0.5$ and (ii) $p_n, p_p \to 0$.  The first case yields $F_{\text{max}} \to 0$ directly. For the second case,  we have
\begin{equation}
    \lim_{p_n, p_p \to 0} F_{\text{max}} = \lim_{p_n, p_p \to 0}[ \frac{1}{4} - \frac{(p_n / p_p)}{(p_n / p_p + 1)^2}].
\end{equation}
Given $p_n < p_p$ and the constraint $\epsilon_s > 0$, it follows that $(1 - 2 p_p)/(1 + 2 p_p) < p_n / p_p < 1$. In the limit $p_p \to 0$,  the term $(1 - 2 p_p)/(1 + 2 p_p)\to 1$, indicating 
$p_n/p_p\to 1$. Thus, $F_{max}=0$ when $p_n, p_p \to 0$.
Consequently, $F < 0$ holds strictly for $0 < \epsilon_s < 1$. Combining with the boundary analysis, $\delta w_{quai}^{{OIE}^2} < \delta w_{quai}^{{Otto}^2} $ is universally satisfied for any measurement error $\epsilon$.

\section{The detailed dynamics of the finite time SOE and OIE}\label{CCCC}

Building upon the quasistatic analyses, we now detail the operational procedure of the finite time Otto cycle:

In the adiabatic compression stroke   $A \to B$ of duration $\tau$,
the system undergoes an unitary evolution governed by the time-dependent Hamiltonian:
\begin{equation}\label{mjkdf}
    H_{ch}(t) = \frac{\omega_{ch}(t)}{2} [\cos(\frac{\pi t}{2\tau})\sigma_x + \sin(\frac{\pi t}{2\tau})\sigma_z],
\end{equation}
where the driving function is $\omega_{ch}(t) = \omega_c (1 - t/\tau) + \omega_h (t/\tau)$. 
The non-commutativity $[H_{ch}(t), H_{ch}(t')] \neq 0$ gives rise to quantum friction.  Then, this friction will induce non-adiabatic transitions between energy levels. Here,
the transition probability is denoted by $\xi = |\langle x_A | U_{ch} | x_B \rangle|^2$ \cite{RJ19,JPS19}, where the unitary evolution operator is expressed as $U_{ch} = \mathcal{T}_> \exp[-i \int_0^{\tau_{ch}} H_{ch}(t) dt]$ with $\mathcal{T}_>$ denoting the time-ordering operator.
Following the two-point measurement (TPM) scheme \cite{lutz07}, the stochastic work $w_{ch,fin}^{Otto}$ is defined as the net change in the system's internal energy during this isolated stroke. 

\begin{table*}
\caption{Finite time SOE trajectory distribution table}\label{tab3}%
\begin{tabular*}{\textwidth}
{@{\extracolsep\fill}ccccccccc}
\toprule
$x_A$ & $x_B$  & $x_C$  & $x_D$ & $w_{ch,fin}^{Otto}$ & $q_{fin}^{Otto}$ & $w_{ch,fin}^{Otto}$& $w_{fin}^{Otto}$ & $p(X_F)$\\
\hline
 + & + & + & + & $\frac{\omega_c-\omega_h}{2}$ & $0$ & $\frac{\omega_h-\omega_c}{2}$ & $0$ & $p_{+}^{A}p_{+}^{C}(1-\xi)^2$ \\
    + & + & + & $-$ & $\frac{\omega_c-\omega_h}{2}$ & $0$ & $\frac{\omega_h+\omega_c}{2}$ & $\omega_c$ & $p_{+}^{A}(1-\xi)p_{+}^{C}\xi$ \\
    + & + & $-$ & + & $\frac{\omega_c-\omega_h}{2}$ & $-\omega_h$ & $-\frac{\omega_h+\omega_c}{2}$ & $-\omega_h$ & $p_{+}^{A}p_{-}^{C}(1-\xi)\xi$ \\
    + & + &$-$ & $-$ & $\frac{\omega_c-\omega_h}{2}$ & $-\omega_h$ & $\frac{\omega_c-\omega_h}{2}$ & $\omega_c-\omega_h$ & $p_{+}^{A}p_{-}^{C}(1-\xi)^2$ \\
    + & $-$ & + & + & $\frac{\omega_c+\omega_h}{2}$ & $\omega_h$ & $\frac{\omega_h-\omega_c}{2}$ & $\omega_h$ & $p_{+}^{A} p_{+}^{C}\xi(1-\xi)$ \\
    + & $-$ & + & $-$ & $\frac{\omega_c+\omega_h}{2}$ & $\omega_h$ & $\frac{\omega_h+\omega_c}{2}$ & $\omega_c+\omega_h$ & $p_{+}^{A} p_{+}^{C}\xi^2$ \\
    + & $-$ & $-$ & + & $\frac{\omega_c+\omega_h}{2}$ & $0$ & $-\frac{\omega_h+\omega_c}{2}$ & $0$ & $p_{+}^{A} p_{-}^{C}\xi^2$ \\
    + & $-$ & $-$ & $-$ & $\frac{\omega_c+\omega_h}{2}$ & $0$ & $\frac{\omega_c-\omega_h}{2}$ & $\omega_c$ & $p_{+}^{A} p_{-}^{C}\xi(1-\xi)$ \\
    $-$ & + & + & + & $-\frac{\omega_c+\omega_h}{2}$ & $0$ & $\frac{\omega_h-\omega_c}{2}$ & $-\omega_c$ & $p_{-}^{A} p_{+}^{C}\xi(1-\xi)$ \\
    $-$ & + & + & $-$ & $-\frac{\omega_c+\omega_h}{2}$ & $0$ & $\frac{\omega_h+\omega_c}{2}$ & $0$ & $p_{-}^{A} p_{+}^{C}\xi^2$ \\
    $-$ & + & $-$ & + & $-\frac{\omega_c+\omega_h}{2}$ & $-\omega_h$ & $-\frac{\omega_h+\omega_c}{2}$ & $-(\omega_h+\omega_c)$ & $p_{-}^{A} p_{-}^{C}\xi^2$ \\
    $-$ & + & $-$ & $-$ & $-\frac{\omega_c+\omega_h}{2}$ & $-\omega_h$ & $\frac{\omega_c-\omega_h}{2}$ & $-\omega_h$ & $p_{-}^{A} p_{-}^{C}\xi(1-\xi)$ \\
   $-$ & $-$ & + & + & $\frac{\omega_h-\omega_c}{2}$ & $\omega_h$ & $\frac{\omega_h-\omega_c}{2}$ & $\omega_h-\omega_c$ & $p_{-}^{A}p_{+}^{C}(1-\xi)^2$ \\
    $-$ & $-$ & + & $-$ & $\frac{\omega_h-\omega_c}{2}$ & $\omega_h$ & $\frac{\omega_h+\omega_c}{2}$ & $\omega_h$ & $p_{-}^{A}p_{+}^{C}\xi (1-\xi)$ \\
    $-$ & $-$ & $-$ & + & $\frac{\omega_h-\omega_c}{2}$ & $0$ & $-\frac{\omega_h+\omega_c}{2}$ & $-\omega_c$ & $p_{-}^{A}p_{-}^{C}\xi(1-\xi)$ \\
   $-$ & $-$ & $-$ & $-$ & $\frac{\omega_h-\omega_c}{2}$ & $0$ & $\frac{\omega_c-\omega_h}{2}$ & $0$ & $p_{-}^{A}p_{-}^{C}(1-\xi)^2$\\
\hline
\end{tabular*}
\end{table*}

In the isochoric heating stroke $B \to C$,  the system interacts with the hot reservoir with a fixed frequency $\omega_h$.
The primary objective of this section is to investigate the interplay between quantum friction and measurement error. 
To isolate the influence of the coherence, we assume that the system fully thermalizes with the hot reservoir at the end of this stroke. 
Consequently, the quantum coherence generated during the  adiabatic compression is completely erased.
We further treat this stroke as instantaneous compared to the driving duration, which can be achieved by high photon density of frequency $\omega_h$.  
The heat $q^{{Otto}}_{fin}$ is determined by the internal energy change via the TPM scheme.
\begin{table*}
\caption{The trajectory distribution table of the finite time OIE under the condition $y=g$.}\label{tab4}%
\begin{tabular*}{\textwidth}
{@{\extracolsep\fill}cccccccccc}
\toprule
$y$ & $x_A$ & $x_B$  & $x_C$  & $x_D$ & $w_{ch,fin}^{OIE}$ & $q_{fin}^{OIE}$ & $w_{ch,fin}^{OIE}$& $w_{fin}^{OIE}$ & $p(X_F,y)$\\
\hline
 $-$ & + & + & + & + & $\omega_c-\omega_h$ & $0$ & $\omega_h-\omega_c$ & $0$ & $p_{+}^A \epsilon p_{+}^{C}(1-\xi)^2$ \\
    $-$ & + & + & + & $-$ & $\omega_c-\omega_h$ & $0$ & $\omega_h$ & $\omega_c$ & $p_{+}^A\epsilon p_{+}^{C}\xi(1-\xi)$ \\
    $-$ & + & + & $-$ & + & $\omega_c-\omega_h$ & $-\omega_h$ & $-\omega_c$ & $-\omega_h$ & $p_{+}^A \epsilon )p_{-}^{C}\xi(1-\xi)$ \\
    $-$ & + & + & $-$ & $-$ & $\omega_c-\omega_h$ & $-\omega_h$ & $0$ & $\omega_c-\omega_h$ & $p_{+}^A \epsilon p_{-}^{C}(1-\xi)^2$ \\
    $-$ & + & $-$ & + & + & $\omega_c$ & $\omega_h$ & $\omega_h-\omega_c$ & $\omega_h$ & $p_{+}^A \epsilon  p_{+}^{C}\xi(1-\xi)$ \\
    $-$ & + & $-$ & + & $-$ & $\omega_c$ & $\omega_h$ & $\omega_h$ & $\omega_c+\omega_h$ & $p_{+}^A \epsilon p_{+}^{C}\xi^2$ \\
    $-$ & + & $-$ & $-$ & + & $\omega_c$ & $0$ & $-\omega_c$ & $0$ & $p_{+}^A\epsilon p_{-}^{C}\xi^2$ \\
    $-$ & + & $-$ & $-$ & $-$ & $\omega_c$ & $0$ & $0$ & $\omega_c$ & $p_{+}^A \epsilon  p_{-}^{C}\xi(1-\xi)$ \\
    $-$ & $-$ & + & + & + & $-\omega_h$ & $0$ & $\omega_h-\omega_c$ & $-\omega_c$ & $p_{-}^A(1-\epsilon) p_{+}^{C}\xi(1-\xi)$ \\
    $-$ & $-$ & + & + & $-$ & $-\omega_h$ & $0$ & $\omega_h$ & $0$ & $p_{-}^A(1-\epsilon) p_{+}^{C}\xi^2$ \\
    $-$ & $-$ & + & $-$ & + & $-\omega_h$ & $-\omega_h$ & $-\omega_c$ & $-\omega_h-\omega_c$ & $p_{-}^A(1-\epsilon) p_{-}^{C}\xi^2$ \\
    $-$ & $-$ & + & $-$ & $-$ & $-\omega_h$ & $-\omega_h$ & $0$ & $-\omega_h$ & $p_{-}^A(1-\epsilon) p_{-}^{C}\xi(1-\xi)$ \\
    $-$ & $-$ & $-$ & + & + & $0$ & $\omega_h$ & $\omega_h-\omega_c$ & $\omega_h-\omega_c$ & $p_{-}^A(1-\epsilon)p_{+}^{C}(1-\xi)^2$ \\
    $-$ & $-$ & $-$ & + & $-$ & $0$ & $\omega_h$ & $\omega_h$ & $\omega_h$ & $p_{-}^A(1-\epsilon)p_{+}^{C}\xi(1-\xi)$ \\
    $-$ & $-$ & $-$ & $-$ & + & $0$ & $0$ & $-\omega_c$ & $-\omega_c$ & $p_{-}^A(1-\epsilon)p_{-}^{C}\xi(1-\xi)$ \\
    $-$ & $-$ & $-$ & $-$ & $-$ & $0$ & $0$ & $0$ & $0$ & $p_{-}^A(1-\epsilon)p_{-}^{C}(1-\xi)^2$ \\

\hline
\end{tabular*}
\end{table*}

\begin{table*}
\caption{The trajectory distribution table of the  finite time OIE under the condition  $y=e$.}\label{tab5}%
\begin{tabular*}{\textwidth}
{@{\extracolsep\fill}ccccc}
\toprule
$y$&$x_A$  & $x_{C^{'}}$  & $q_{fin}^{OIE}$ &  $p(X_F)$\\
\hline
 $+$&$+$ & $+$ & $0$ & $p_+^A(1-\epsilon)p_+^{C^{'}}$ \\
$+$&$+$ & $-$ & $-w_c$ & $p_+^A(1-\epsilon)p_-^{C^{'}}$ \\
 $+$&$-$ & $+$ & $w_c$ & $p_-^A\epsilon p_+^{C^{'}}$ \\
 $+$&$-$ & $-$ & $0$ & $p_-^A\epsilon p_-^{C^{'}}$\\
\hline
\end{tabular*}
\end{table*}
In the adiabatic expansion stroke $C \to D$, 
the system is again isolated from the reservoirs. The driving protocol is the time-reversal of the compression stroke  $\omega_{hc}(t) = \omega_{ch}(\tau - t)$, which ensures that the transition probability  $\xi$ remains identical for both adiabatic stroke.
The internal energy change is fully converted into work, and the stochastic work output $w_{hc,fin}^{Otto}$ is also determined via TPM.

In the isochoric cooling ($D \to A$), the cycle is closed at frequency $\omega_c$ as the system relaxes back to its initial equilibrium by interacting with the cold reservoir.

The stochastic trajectories $X_F$ of the system  over a complete cycle, along with their associated work stochastic $w_{fin}^{Otto}=w_{ch,fin}^{Otto}+w_{hc,fin}^{Otto}$, absorbed heat $q_{fin}^{Otto}$ , and the probabilities of $X_F$, are summarized in Tab.~\ref{tab3}.
According to Tab. \ref {tab3}, the average work $\langle w_{fin}^{Otto}\rangle$, the work fluctuations $\delta w_{fin}^{Otto^2}$, and the average heat $\langle q_{fin}^{Otto}\rangle$ in the finite time SOE can  be  determined 
\begin{eqnarray}
\langle w_{fin}^{Otto}\rangle&=&\int p(X_F)w_{fin}^{Otto}(X_F)dX_F\nonumber\\
   &=&(\omega_h-\omega_c)(p_-^{A} p_+^{C} - 
p_+^{A} p_-^{C}  )\nonumber\\
   &+&  [(p_+^{A}- p_-^{A}) \omega_c + (p_+^{C} - p_-^{C} )
 \omega_h] \xi,\label{waf1}
 \end{eqnarray}
 \begin{eqnarray}
    \delta w_{fin}^{Otto^2}&=&\int p(X_F)(w_{fin}^{Otto}(X_F))^2dX_F
    -\langle w_{fin}^{Otto}\rangle^2\nonumber\\
   &=& \omega_h^2[\frac{1}{2}-\frac{(p_+^A-p_-^A)^2(1-2\xi)^2+(p_+^C-p_-^C)^2}{4}]\nonumber\\
   &+&  \omega_c^2[\frac{1}{2}-\frac{(p_+^A-p_-^A)^2+(p_+^C-p_-^C)^2(1-2\xi)^2}{4}]\nonumber\\
&+&\omega_c\omega_h[1-\frac{(p_+^A-p_-^A)^2+(p_+^C-p_-^C)^2}{2}],\label{jkl1} 
\end{eqnarray}
 \begin{eqnarray}
\langle q_{fin}^{Otto}\rangle&=&\int p(X_F)q_{finite}^{Otto}(X_F)dX_F\nonumber\\
   &=& \omega_h[p_-^{A} p_+^{C}- p_+^{A} p_-^{C} + (p_+^{C} - p_-^{C})\xi].\label{adc1}
\end{eqnarray}

Integrating the feedback protocol into the  finite time SOE, the OIE operation is determined by  measurement outcomes 
$y$ and the stochastic trajectories  $X_F $. The associated work and heat contributions for each possible process are summarized in Tab. ~\ref{tab4} and Tab. \ref{tab5} . From these distributions, the average work, the work fluctuations and the average heat absorption  of the finite time OIE can be derived

\begin{eqnarray}
    \langle w^{OIE}_{fin}\rangle&=&\int p(X_F, y=g)w_{fin}^{OIE}(X_F,y=g)d X_F\nonumber\\
    &=&p_-^{A} p_+^{C} (\omega_h-\omega_c) + p_-^{A} [(p_+^{C} -p_-^{C})\omega_c-\omega_h]\xi\nonumber\\
    &+&\{(p_-^{A} p_+^{C}+p_+^{A} p_-^{C}) (\omega_c - \omega_h)\nonumber\\
    &+&[\omega_h+(p_+^{A}-p_-^{A})(p_+^{C} - p_-^{C})\omega_c]\xi\}\epsilon,
    \end{eqnarray}

    \begin{eqnarray}
    \delta w^{OIE^2}_{fin}&=& \int p(X_F, y=g)[w_{fin}^{OIE}(X_F,y=g)]^2d X_F\nonumber\\
    &-&\langle w^{OIE}_{fin}\rangle^2\nonumber\\
   &=& (1-\xi)^2 (\omega_c - \omega_h)^2  [(1-\epsilon) p_-^A p_+^C + \epsilon p_+^A p_-^C]\nonumber \\
& +& \xi^2 (\omega_c + \omega_h)^2 [ \epsilon p_+^A p_+^C + (1-\epsilon) p_-^A p_-^C] \nonumber\\
& +& \xi(1-\xi) (\omega_c^2 + \omega_h^2) [ \epsilon p_+^A + (1-\epsilon) p_-^A ] \nonumber\\
& -& \{ (1-\xi)(\omega_c - \omega_h) [ \epsilon p_+^A p_-^C - (1-\epsilon) p_-^A p_+^C ]\nonumber\\
 &+& \xi(\omega_c + \omega_h) [\epsilon p_+^A p_+^C - (1-\epsilon) p_-^A p_-^C ] \}^2, 
 \end{eqnarray}
 \begin{eqnarray}
\langle q_{fin}^{OIE}\rangle&=&\int p(X_F,y)q_{finite}^{OIE}(X_F,y)dX_Fdy\nonumber\\
&=&\omega_h[(1-\epsilon) p^{A}_- p^{C}_+-\epsilon p^{A}_+ p^{C}_-+(\epsilon-p^{A}_-)\xi]\nonumber\\
&+&\omega_c[p^{A}_- \epsilon p_+^{C^{'}} - p^{A}_+ (1 - \epsilon) p_-^{C^{'}}].
\end{eqnarray}

\begin{figure*}
\begin{overpic}[width=0.9\textwidth]{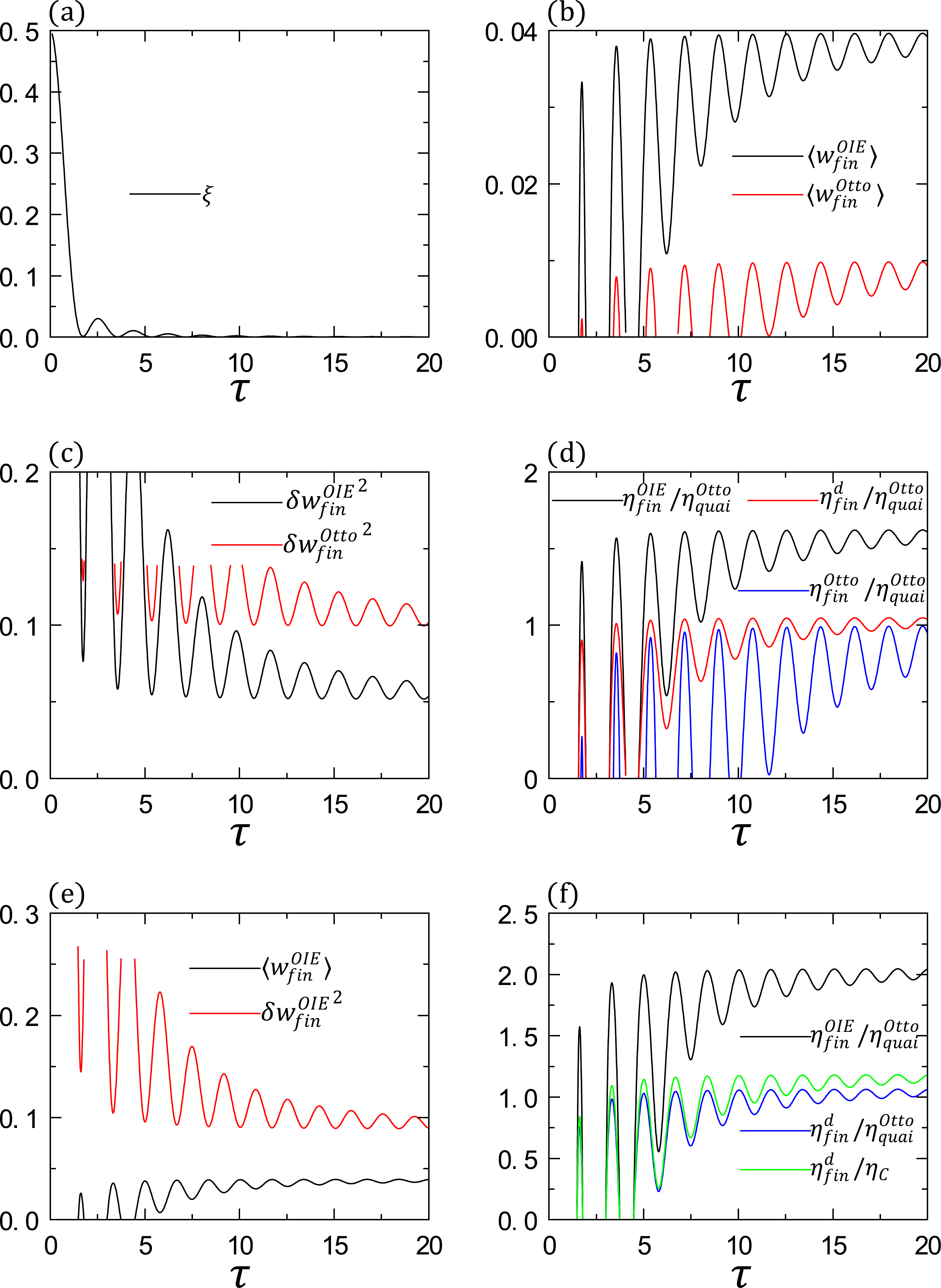}
\end{overpic}
\caption{The performance parameters  of the finite-time OIE as  functions of the driving time $\tau$. (b)-(d) correspond to $\omega_h = 4$, while  (e) and (f) correspond to $\omega_h = 4.5$ with measurement error  fixed at $\epsilon = 0.15$. Other parameters are the same as in Fig.~\ref{werr}.}\label{knj}
\end{figure*}

In Fig. \ref{knj}(a), the transition probability $\xi$ 
exhibits an oscillatory decay with increasing driving time $\tau$, eventually vanishing as the system reaches the quantum adiabatic limit. Consequently, the performance parameters (the work, work fluctuations and the efficiency) in Figs. \ref{knj}(b)-(f) asymptotically approach their respective quasistatic values. Notably, the curves are interrupted in the short-time regime. These discontinuities arise due to that high transition probability $\xi$ causes large friction  work, which  renders positive work extraction impossible. 
 
A comparative analysis across Figs.~\ref{knj}(b)-(d) confirms that the OIE consistently outperforms the SOE. Specifically, the OIE yields higher work output, suppressed work fluctuations, and enhanced efficiency—even when the energetic cost of the demon is fully accounted for. This demonstrates that the  advantages of the OIE are not limited to the quasistatic regime but persist  under finite time operation.

To further elucidate the performance trends, we present the work, the work fluctuations, and the efficiency  in Figs. \ref{knj}(e) and (f) under the regime where the SOE operates as a refrigerator. A striking feature observed in panel (f) is that the finite time OIE efficiency with  the demon's cost can surpass both the quasi-static Otto limit and the standard Carnot bound. This regime is inaccessible to the  finite time SOE  in which $\eta_{fin}^{Otto}<\eta_{quai}^{Otto}<\eta_C$ holds strictly. 

\newpage
\nocite{*}
\bibliography{apssamp}

\end{document}